%% file: starkman_etal_2023_diffusion_distortion.tex
\documentclass[fleqn,usenatbib]{mnras}
\usepackage{showyourwork}

\usepackage{hyperref}
\usepackage{amsmath, amssymb}
\usepackage{newtxmath}
\usepackage{enumitem}
\usepackage{graphicx}
\usepackage{flushend}
\usepackage{cuted}
\usepackage{widetext}

\usepackage{tikz}
\usepackage{mathdots}
\usepackage{cancel}
\usepackage{color}
\usepackage{siunitx}
\usepackage{array}
\usepackage{multirow}
\usepackage{gensymb}
\usepackage{tabularx}
\usepackage{extarrows}
\usepackage{booktabs}
\usetikzlibrary{fadings}
\usetikzlibrary{patterns}
\usetikzlibrary{shadows.blur}
\usetikzlibrary{shapes}

\tikzset{every picture/.style={line width=0.75pt}} %set default line width to 0.75pt

\input{preamble.tex}

\DeclareRobustCommand{\VAN}[3]{#2} \let\VANthebibliography\thebibliography
\def\thebibliography{\DeclareRobustCommand{\VAN}[3]{##3}\VANthebibliography}

\title[Angular Correlations of CMB Spectrum Distortions from Photon Diffusion]{
Angular Correlations of Cosmic Microwave Background Spectrum Distortions from
Photon Diffusion}

\author[N. Starkman et al.]{ Nathaniel Starkman$^{1}$\thanks{E-mail:n.starkman@mail.utoronto.ca}, Glenn Starkman$^{2,3}$, Arthur Kosowsky$^{4}$ \\
$^{1}$David A. Dunlap Department of Astronomy and Astrophysics, University of Toronto, 50 St. George Street, Toronto, Ontario, M5S 3H4, Canada\\
$^{2}$Physics Department/CERCA/ISO Case Western Reserve University Cleveland, Ohio 44106-7079, USA\\
$^{3}$Astrophysics Group \& Imperial Centre for Inference and Cosmology, Department of Physics, Imperial College London, Blackett Laboratory, Prince Consort Road, London SW7 2AZ, United Kingdom\\
$^{4}$Department of Physics and Astronomy, University of Pittsburgh, Pittsburgh, PA 15260, USA\\
}

\date{Accepted XXX.  Received YYY; in original form ZZZ}

\pubyear{2023}

\begin{document}
\pagerange{\pageref{firstpage}--\pageref{lastpage}}
\maketitle

\begin{abstract}

    During cosmic recombination, charged particles bind into neutral atoms and
    the mean free path of photons rapidly increases, resulting in the familiar
    diffusion damping of primordial radiation temperature variations.  An
    additional effect is a small photon spectrum distortion, because photons
    arriving from a particular sky direction were originally in thermal
    equilibrium at various spatial locations with different temperatures; the
    combination of these different blackbody temperature distributions results
    in a spectrum with a Compton $y$-distortion.  Using the approximation that
    photons had zero mean free path prior to their second-to-last scattering, we
    derive an expression for the resulting $y$-distortion, and compute the
    angular correlation function of the diffusion $y$-distortion and its
    cross-correlation with the square of the photon temperature fluctuation.
    Detection of the cross-correlation is within reach of existing
    arcminute-resolution microwave background experiments such as the Atacama
    Cosmology Telescope and the South Pole Telescope.

\end{abstract}


%%%%%%%%%%%%%%%%%%%%%%%%%%%%%%%%%%%%%%%%%%%%%%%%%%%%%%%%%%%%%%%%%%%%%%%%%%%%%%%

%%%%%%%%%%%%%%%%%%%%%%%%%%%%%%%% BODY OF PAPER %%%%%%%%%%%%%%%%%%%%%%%%%%%%%%%%

\label{firstpage}

\section{Introduction} \label{sec:intro}

    The average of two or more blackbody spectra at different temperatures is
    not a perfect blackbody.  The Cosmic Microwave Background (CMB) photons
    arriving from a particular line of sight (LOS) scattered multiple times
    during recombination before their last-scattering into that LOS.  Because
    CMB photons at the epoch of last scattering have energies that are far
    smaller than the electron rest mass, the fractional energy change of a
    scattered photon is small (of order $E_\gamma / m_e c^2$ by the
    Klein-Nishina formula \citep[as translated in
    \cite{KleinNishina1994}]{KleinNishina1929} and of order $v_e^2/c^2$ for
    effects due to the electron's velocity).  Therefore, while electron
    thermalization with the CMB continues through recombination, as does photon
    isotropization, to a good approximation photon last-scattering does not
    change the photon's energy but only its propagation direction.

    If the photons of the CMB came to us along each LOS unscattered from a
    ``surface of last thermal emission,'' then they would be drawn from the
    thermal-equilibrium blackbody distribution in their local neighborhood of
    emission. However, there is no ``surface of last thermal emission.''  
    Instead, the CMB photons are scattered with nearly no change in energy into
    the LOS at their point of last scattering.  The photon energies are
    therefore drawn, approximately, from the black-body distributions of the
    neighborhoods from which the photons originated.  The energy distribution of
    the photons arriving along each LOS is approximately the average of many
    blackbodies, each with the temperature of a different neighborhood of photon
    emission.

    A simple approximation to this averaging effect is to assume that each CMB
    photon scatters exactly once after its emission from a blackbody, or
    equivalently that the photon mean-free-path is negligible prior to its
    second-to-last scattering (2LS).  We therefore take the point of emission,
    whose temperature distribution the photon samples, to be the point of
    second-last scattering.  In this approximation, we provide an analytic
    expression for the contribution of any given Fourier mode of adiabatic
    perturbation to the photon spectrum from a given sky direction.  The net
    effect of all such perturbation modes is an integral over the contribution
    of each mode, which can be evaluated numerically.  The first-order effect
    (in the difference between the temperatures in the second-last-scattering
    neighborhood and the global average temperature at that cosmic time) is a
    blackbody photon distribution with a temperature averaged over the region of
    second-to-last scatterings; this averaging approximates the mechanism behind
    the familiar diffusion damping of temperature anisotropies on small angular
    scales.  The second-order effect is a  $y$-distortion of the blackbody
    distribution first described (but not quantified) in \citep{Zeldovich+1972},
    and recalled in \citep{ChlubaSunyaev2004}.  More details were provided in
    \citep{Khatri+2012}, but again without specific predictions.  A calculation
    of the mean expected distortion may be among the effects included in
    \citep{Chluba+2012}.  Similarly, it is described in
    \citep{SunyaevKhatri2013,Chluba2016}.  \citet{Stebbins:2007ve} (unpublished)
    states that scattering of temperature anisotropies inevitably leads to
    spectral distortions of the CMB, but argues that pre-recombination
    distortions will be overwhelmed by foreground Sunyaev-Zeldovic (SZ)
    distortions.  The authors of \cite{Pitrou:2009bc} similarly calculate the
    angular spectrum of $y$ distortions, and find that it is dominated by
    reionization-era effects.  Stebbins does suggest \citep{Stebbins:2007ve}
    checking whether in \LCDM{} cosmologies one can remove the thermal SZ
    contamination by correlating anisotropies in the spectral distortion with
    anisotropies in temperature. We do so in this paper.

    We calculate this effect in \LCDM, showing that the $y$-distortion angular
    correlation function is at a level that is potentially measurable in future
    experiments, but confirm that it is small compared to other expected signals
    \citep{ChlubaSunyaev2004}.  However, we also calculate the angular
    cross-correlation function of this y-distortion and the square of the
    primordial temperature fluctuations.  The cross-correlation is likely
    detectable in current experiments, including the Atacama Cosmology Telescope
    (ACT) \citep{Coulton+2023} and the South Pole Telescope (SPT)
    \citep{SPTymap}, and especially in anticipated experiments like the Simons
    Observatory \citep{Galitzki+2018} and CMB-S4 \citep{Abitbol+2017}.  The
    authors of \citet{Chluba+2022} and \citet{Kite+2022} address related
    questions, including cross-correlations, although not specifically this
    effect.

    The mixing of blackbody photon distributions is distinct from the averaging
    of temperatures over different lines of sight, whether through the finite
    width of telescope beams \cite[e.g.][]{ChlubaSunyaev2004} or through the
    angular integration inherent in the calculation of the coefficients of
    spherical harmonics \cite[e.g.][]{Chluba+2017,Lucca+2020}.  Each of these
    effects also creates small $y$-distortions in measured signals through
    temperature averaging, but the effect discussed here is due to physical
    processes during the epoch of last scattering and thus carries information
    about the universe rather than about a given measurement.  A different
    application of spectral distortions due to blackbody mixinga through
    scattering was considered in \cite{CaldwellStebbins2008} as a way to
    constrain inhomogeneities at reionization.

    In \autoref{sec:scattering-distribution}, we calculate the distribution of
    second-last-scattering locations.  In \autoref{sec:calculating_the_signal},
    we calculate the expected $y$-distortion $\Yd$ of the CMB spectrum due to
    this diffusive averaging.  In \autoref{sec:correlation_functions}, we
    calculate the expected angular correlation function of $\Yd$, as well as the
    expected cross-correlation between $\Yd$ and the square of the observed
    temperature fluctuation $(\DeltaT)^2$.  We discuss the detectability of this
    signal in \autoref{sec:detectability}, and conclude in
    \autoref{sec:conclusion}.

    \begin{figure*}
        \include{graphic.tex}
        \caption{%
            The coordinate system. The observer is at a location $z_0 > 0$,
            looking along a line of sight $\hat{n}$. From the observer's
            position are defined the cylindrical coordinates ($s_{\perp}, \phi,
            s_{||}$). A photon's $\phi$ is its azimuthal angle of last
            scattering into the line of sight. The last scatter occurs at $z_1$
            at a comoving distance $r_{01}$ from $z_0$. From the
            2nd-last-scatter at $z_2$, the photon enters the last scattering
            location with inclination angle $\theta$.
        }
        \label{fig:coordinate_system}
    \end{figure*}

% section introduction (end)

\section{Second-to-Last Scattering Distribution}
\label{sec:scattering-distribution}

    Photons arriving at a common point of last scattering $\mbf{x_1}$ come from
    a surrounding neighborhood of prior locations $\mbf{x_2}$, where we use the
    subscript to label the scattering number, counting back in time from
    observation at scattering ``0''.  Arriving at $\mbf{x_1}$, the photons
    scatter into the LOS and their combined distribution is no longer that of a
    perfect blackbody; it is an average of blackbodies.  The resulting
    distortion of the blackbody spectrum retains information about the
    distribution of temperatures in the second-last-scattering neighborhood that
    goes beyond the intensity of the photon emissions from that LOS.

    \begin{figure}
        \script{p2ls/plot_p2ls.py}
        \centering
        \includegraphics[width=1.05\columnwidth,height=2in]{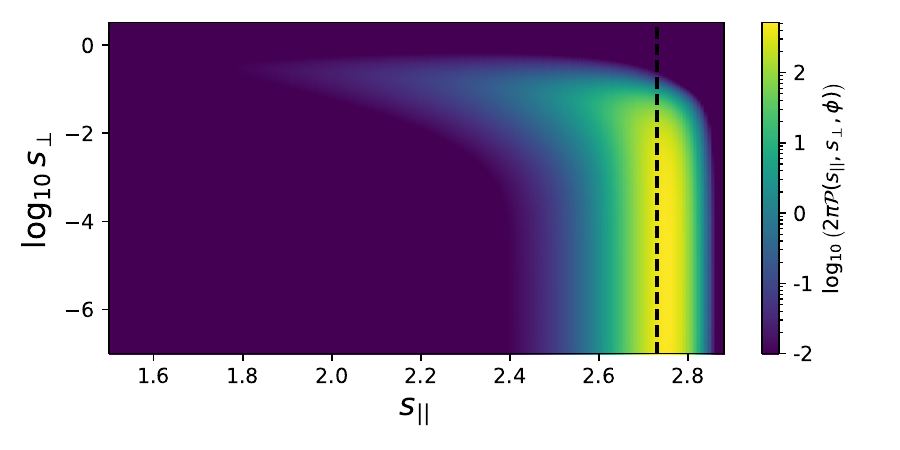}
        \vspace{-20pt}
        \caption{%
            $\mcal{P}(\spll, \sprp, \phi=0)$ over a range of $\spll, \sprp$,
            colored by $\log_{10}\!\left(2\pi\mcal{P}\right)$.  $\mcal{P}$ peaks
            at $\spll\simeq 2.76$ -- just slightly higher than $\sreco=2.73$
            (dashed line) -- and at $\sprp \lesssim 10^{-3}$ (below which it is
            approximately constant).  When sampling from $\mcal{P}$ each decade
            in $\sprp$ contributes approximately one-tenth the samples as the
            previous decade; thus we display only  $\sprp \geq 10^{-7}$.  The
            other boundaries of the domain of $\mcal{P}$ are taken at sufficient
            distance from the higher probability regions of $\mcal{P}$ that
            enlarging the domain does not impact the results.%
        }
        \label{fig:P2LS}
    \end{figure}

    To calculate the observational effects of this blackbody averaging, we must
    first calculate the probability density $\mcal{P}(\mbf{x_2},\mbf{x_1})$ that
    a photon arriving at the observer from a last-scattering location
    $\mbf{x_1}$ had its second-last scattering at $\mbf{x_2}$.  While we refer
    to our location as ``the observer,'' in order to remove the physically
    distinct effects of propagation through the low-redshift universe, we
    consider a reference point $\mbf{x_0}$ at co-moving distance $r_0$ from us
    along the LOS, which is the line between the observer and $\mbf{x_1}$.  We
    take $\mbf{x_0}$ to be at a redshift $z_0=100$, well after recombination and
    last scattering (at $z\simeq10^3$) but well before the onset of cosmic
    acceleration.  The comoving displacement of the j-th-last scatter from the
    i-th-last-scattering location is $\mbf{x_{ij}}\equiv\mbf{x_j}-\mbf{x_i}$.
    We also write $r_{ij}$ for $\norm{\mbf{x_{ij}}}$. The details of the
    coordinates are shown in \autoref{fig:coordinate_system}.

    To leading order (in fluctuations around the homogeneous background metric
    and stress-energy tensor), the probability distribution $\mcal{P}(\mbf{x_2},
    \mbf{x_1})$ is rotationally symmetric about the line-of-sight axis.
    $\mcal{P}(\mbf{x_2},\mbf{x_1})$ can thus be written in terms of:

    \begin{enumerate}[leftmargin=*]
        \item
            $P_1(r_{01})\,\dif{r_{01}}$, the probability that last scattering
            occurs at a comoving distance from $\mbf{x_0}$ between $r_{01}$ and
            $r_{01}+\dif{r_{01}}$ along the LOS
            \begin{equation}  \label{eq:P1r1}
                P_1(r_{01}) = \GgamBar(r_0 + r_{01},r_0)\,,
            \end{equation}
            where $\GgamBar(r_a,r_b)$ is the visibility function for Thomson
            scattering between comoving radii $r_a$ and $r_b\geq r_a$ (see
            \appref{app:visibility_function});
        \item
            $P_\theta(\theta)\,\dif{\theta}$, the probability that the last
            scattering is through an angle between $\theta$ and
            $\theta+\dif{\theta}$
            \begin{equation} \label{eq:Ptheta}
                P_\theta(\theta)
                    = \frac{\dif{\sigma_T}}{\dif{\theta}}
                    = \frac{3}{8}\sin\theta\left(1+\cos^2\theta\right)
                      \,\,,\ \theta \in [0, \pi]\,;
            \end{equation}
        \item
            $P_\phi(\phi)$, the probability distribution of the last scattering
            azimuth
            \begin{equation} \label{eq:Pphi}
               P_\phi(\phi) = \frac{1}{2\pi} \,\,,\ \phi \in [0, 2\pi)\,;
            \end{equation}
        \item
            $P_2(r_{12}|r_{01}) \, dr_{12}$, the probability of the 2nd-to-last
            scatter taking place at a comoving distance between $r_{12}$ and
            $r_{12} + dr_{12}$ from the last scattering at comoving distance
            $r_{01}$ from $\mbf{x_0}$
            \begin{equation} \label{eq:P2r12_given_r1}
                P_2(r_{12}|r_{01}) = \GgamBar(r_0 + r_{01}+r_{12},r_0 + r_{01})\,.
            \end{equation}
    \end{enumerate}

    The overbar on $\GgamBar$ in \autoref{eq:P1r1} and \ref{eq:P2r12_given_r1}
    indicates that we are considering $\GgamBar$ to be a function only of
    redshift, i.e. a homogeneous function of location on a given redshift slice.
    The arguments of $\GgamBar$ are cosmological epochs, not positions; thus in
    \autoref{eq:P2r12_given_r1}, $r_{01}+r_{12} \neq r_{02}$.  From
    \autoref{eq:P1r1}, it is clear that $\GgamBar(r_a,r_b)$ is normalized such
    that $\int\dif{r_a} \GgamBar(r_a,r_b)=1$.  We note that $P_2$ is independent
    of scattering angle $\theta$ (to leading order).

    Multiplying these four probabilities -- $P_1$, $P_\theta$, $P_\phi$, and
    $P_2$ -- gives the probability density $\mcal{P}$ for the last scattering to
    take place at comoving distance $r_{01}$ from the observer and the
    second-last scattering to take place at comoving distance $r_{12}$ from the
    last scattering, with scattering angle $\theta$ and azimuthal angle $\phi$.

    We prefer to express $\mcal{P}$ in terms of new dimensionless comoving
    coordinates $\spll$ and $\sprp$ measured respectively along the LOS (in the
    $\nhat$ direction) and perpendicular to it, instead of $r_{12}$ and
    $\theta$.%
    Thus
    \begin{equation} \label{eq:xofs}
        \mbf{x}(\mbf{s};\nhat) = r_0 \nhat + \Leq\mbf{s}\,,
    \end{equation}
    where $\Leq$ is defined in \appref{app:coordinate_systems}.  We define the
    vector $\mbf{s}$, with origin at $\mbf{r_0}$, i.e. along the LOS at redshift
    $z_0$.  $\mbf{s}$ is directed away from the reference point and away from
    the observer; it can be decomposed as
    \begin{equation}
        \mbf{s} = \spll \nhat + \mbf{s}_\perp \,.
    \end{equation}
    $\mbf{s}_\perp$ is a two-vector of magnitude $\sprp$ in the plane
    perpendicular to $\nhat$.  The direction of $\mbf{s}_\perp$ is specified by
    $\phi$.

    \begin{equation} \label{eq:s_definition}
        \spll = \frac{r_{01}}{\Leq} + \frac{r_{12}}{\Leq}\cos\theta \quad \mathrm{and}\quad
        \sprp = \frac{r_{12}}{\Leq}\sin\theta.
    \end{equation}
    We can take the new set of variables to be $r_{01}$, $\sprp$, $\spll$, and
    $\phi$, leading to a Jacobian factor
    \begin{equation} \label{eq:r2s_jacobian}
        J = \frac{\Leq^3}{r_{12}}
          = \frac{\Leq^3}{\left((\Leq \spll-r_{01})^2 + \Leq^2\sprp^2\right)^{1/2}}\,.
    \end{equation}
    In terms of these new variables
    \begin{equation}
        \mcal{P}_\theta(\theta)
          = \left( \frac{3\Leq\sprp}{8} \right)
            \frac{2(\Leq \spll-r_{01})^2 + \Leq^2\sprp^2}
                 {\left((\Leq \spll-r_{01})^2 + \Leq^2\sprp^2\right)^{3/2}}
    \end{equation}

    The probability density $\mcal{P}(\spll,\sprp,\phi)$ is then

    \begin{widetext}
    \begin{align} \label{eq:Pspllsprpphi}
        \mcal{P}(\spll,&\sprp,\phi)
            = \int\!\!\!\dif{r_{01}} \; \mcal{P}(r_{01},r_{12},\theta,\phi) J 
             = \int\!\!\!\dif{r_{01}} \; P_2(r_{12}|r_{01}) \, P_1(r_{01}) \, P_\theta(\theta) \, P_\phi(\phi) \, J
            \\
            &= \frac{\Leq^2}{2\pi} \!\!
               \int\!\!\!\dif{r_{01}} \;
                \GgamBar(r_0+r_{01},r_0) \,
                \GgamBar\left( r_0+r_{01}+\sqrt{(\Leq \spll-r_0+r_{01})^2+\Leq^2\sprp^2}, r_0+r_{01} \right)
                \left( \frac{3\Leq \sprp}{8} \right)
                \frac{2(\Leq \spll-r_{01})^2 + \Leq^2\sprp^2}
                    {\left((\Leq \spll-r_{01})^2+\Leq^2\sprp^2\right)^2} \,, \nonumber
    \end{align}
    \end{widetext}
    where the variables are $\sprp$ and $\spll$, not $\sprp^2$ and
    $\spll$, thus
    \begin{equation} \label{eq:Pnormalization}
        1 = \int_0^{\infty}\!\!\!\dif{\sprp} \dif{\spll} \int_0^{2\pi}\!\!\!\dif{\phi} \; \mcal{P}(\spll,\sprp,\phi)
          = \int\!\!\dif[3]{s}\; \mcal{P}(\spll,\sprp,\phi) \,. \nonumber
    \end{equation}

    The numerical evaluation of $\mcal{P}(\spll,\sprp,\phi)$ is described in
    \appref{app:Pspllsprp}, and the result is presented in \autoref{fig:P2LS}.
    We observe that $\mcal{P}$ is sharply peaked close to the position of
    recombination, i.e at $\spll \simeq \sreco$, and with $\sprp \lesssim
    10^{-3}$.

% section second_last_scattering_distribution (end)

\section{Calculating the Signal} \label{sec:calculating_the_signal}

    Consider a sum of blackbodies, each with photon occupation number
    $n_{Pl}(\chi)=(e^\chi-1)^{-1}$ for $\chi\equiv h\nu/kT_i$.  To third order
    in the temperature fluctuations $\epsilon_i \equiv
    \dfrac{T_i-\bar{T}}{\bar{T}}$:
    \begin{align} \label{eqn:sumofbb}
        \sum_i & a_i n_{Pl}(\chi,\!T_i)
            =   n_{Pl}(\chi,\bar{T}) \!\! \sum_i\!a_i
              + n_{Pl}^2(\chi,\bar{T}) \chi e^\chi \! \sum_i \! a_i \epsilon_i 
            \\ & \phantom{=}
              + \frac{1}{2} n_{Pl}^3(\chi,\bar{T}) \left((\chi^2\!\!-\!2\chi) \Exp{2\chi} \!+\! (\chi^2 \!\!+\! 2\chi)e^\chi\right) \!\sum_i\!a_i\epsilon_i^2
              + \mcal{O}(\epsilon_i^3)  \nonumber
    \end{align}
    We can choose to normalize $a_i$ so that $\sum_i a_i=1$.  Taking
    $\bar{T}\equiv\sum_i a_i T_i$ as one would expect, then $\sum_i
    a_i\epsilon_i=0$.

    It is conventional \citep{Chluba+2012} and convenient to reorganize
    \autoref{eqn:sumofbb}, writing:
    \begin{align} \label{eqn:sumofbb_reorg}
        \sum_i a_i n_{Pl}(\chi,\!T_i)
            &=   n_{Pl}(\chi,\hat{T})
               + \left[\sum_i \! a_i (\delta_i - \delta_i^2)\right] n_{Pl}^2(\chi,\hat{T}) \chi e^\chi \nonumber
            \\ &\!\!\!\!\!\!\!\!\!\!
               + \frac{1}{2}\!\left[\sum_i \! a_i \delta_i^2\right] n_{Pl}^3(\nu,\hat{T}) \chi^2
               e^\chi (e^\chi +1)
               + \mcal{O}(\delta^3)
    \end{align}
    where 
    \begin{equation}
        \delta_i\equiv\frac{T_i-\hat{T}}{\hat{T}}.
    \end{equation}
    We are free to choose $\hat{T}$ and do so such that  
    \begin{equation} \label{eqn:Ttilde_def}
        \sum_i a_i (\delta_i - \delta_i^2) = 0\,.
    \end{equation}
    This choice of $\hat{T}$ means \autoref{eqn:sumofbb_reorg} is a pure
    $y$-distortion:
    \begin{align}
        \sum_i a_i & n_{Pl}(\chi,T_i) =
            n_{Pl}(\chi,\hat{T}) \\ &\phantom{=} 
            + \frac{1}{2}\left[\sum_i a_i \delta_i^2\right]n_{Pl}^3(\chi,\hat{T}) \chi^2 e^\chi (e^\chi +1)
            + \mcal{O}(\delta^3)\,. \nonumber
    \end{align}

    Clearly $\hat{T}\neq\bar{T}$, but how different are they, or more
    importantly how different are $\bar{\delta^2}\equiv\sum_i a_i \delta_i^2$
    and $\bar{\epsilon^2}\equiv\sum_i a_i \epsilon_i^2$? We can solve the
    quadratic  \autoref{eqn:Ttilde_def} for $\hat{T}$ and expand $\hat{T}$
    around $\bar{T}$ to find
    \begin{equation}
        \hat{T} = \bar{T} \left[
                  1
                - \frac{1}{2}\bar{\epsilon^2}
                + {\cal{O}}(\bar{\epsilon}^4)
            \right] \,.
    \end{equation}
    It is straightforward to show that
    \begin{equation}
        \delta_i = \epsilon_i - (1+\epsilon_i) \bar{\epsilon^2} + {\cal{O}}(\epsilon^4)\,.
    \end{equation}
    and thus
    \begin{equation}
        \sum_i a_i\delta_i^2 = \sum_i a_i\epsilon_i^2 + {\cal{O}}(\epsilon^4)\,.
    \end{equation}
    For small fluctuations, we can therefore replace $\hat{T}$ by $\bar{T}$, and
    $\delta_i$ by $\epsilon_i$ in the calculation of the signal, which proves
    considerably simpler.

    In the CMB, photons were scattered into the LOS at last-scattering from
    different locations of second-last scattering.  We take the photon
    distribution originating from each 2LS point to be a blackbody of the
    temperature characteristic of the plasma there.  In this approximation, the
    resulting observable photon distribution is the weighted average of the 2LS
    blackbodies at all the accessible 2LS points.  We must therefore account for
    the temperature at each location $\mbf{x}(\mbf{s},\nhat)$ that can scatter
    into the LOS, and we must weight the sum by the probability
    $\mcal{P}(\mbf{s})$ that the photons we see come from that
    location.\footnotemark

    The temperature at $\mbf{x}(\mbf{s}, \nhat)$ is given by
    \begin{align} \label{eqn:Tofx}
        T(\mbf{x}(\mbf{s}, \nhat))
            &=    
                  T_0
                + \int \!\!\!\dif[3]{k} \, A^{\!\mcal{T}\!\!}(\mbf{k}) \, \Exp{i \mbf{k} \cdot \mbf{x}(\mbf{s}, \nhat)} \\
            &\equiv
                  T_0
                + \DeltaT(\mbf{x}(\mbf{s}, \nhat))\,, \nonumber
    \end{align}
    where $T_0$ is the mean CMB temperature.  $A^{\!\mcal{T}\!\!}(\mbf{k})$
    includes the effect of the transfer function $\mcal{T}(\norm{\mbf{k}})$ on
    the primordial amplitude $A(\mbf{k})$ of the Fourier mode of the primordial
    curvature fluctuation with wave vector $\mbf{k}$:
    \begin{equation}
        A^{\!\mcal{T}\!\!}(\mbf{k})\equiv \mcal{T}(\norm{\mbf{k}}) A(\mbf{k}) \,.
    \end{equation}
    For simplicity, we will include only the dominant Sachs-Wolfe term in the
    transfer function, so we can take $\mcal{T}$ and thus $A^{\!\mcal{T}\!\!}$
    to be independent of $\hat{n}$.

    The appropriate weighted mean temperature of the photons scattered into the
    LOS in direction $\nhat$ is
    \begin{align}
        \bar{T}(\nhat)
            &=
                \int\!\!\!\dif[3]{s}\, \mcal{P}(\mbf{s})\, \left(
                      T_0
                    + \int \dif[3]{k} A^{\!\mcal{T}}(\mbf{k}) \Exp{i\mbf{k}\cdot\mbf{x}(\mbf{s},\nhat)}
                \right)
            \\
            &=
                T_0 + \Delta \bar{T}(\nhat) \,.
    \end{align}

    The $y$-distortion signal is 
    \begin{align}
        \Yd(\nhat)
            &=  \frac{1}{2}
                \int\!\!\!\dif{\mbf{s}}\;
                    \mcal{P}(\mbf{s})
                    \left(\frac{T(\mbf{x}(\mbf{s};\nhat))-\bar{T}(\nhat)}{\bar{T}(\nhat)}\right)^2
            \\
            &\simeq
                \frac{1}{2T_0^2}
                \int\!\!\!\dif{\mbf{s}}\;
                    \mcal{P}(\mbf{s})
                    \left( \DeltaT(\mbf{x}(\mbf{s};\nhat))- \Delta \bar{T}(\nhat) \right)^2
            \\
            &=
                \frac{1}{2T_0^2}
                \bigg[
                    \int\!\!\!\dif{\mbf{s}}\; \mcal{P}(\mbf{s}) \left( \DeltaT(\mbf{x}(\mbf{s};\mbf{\nhat}))\right)^2 & \label{eq:signal_realspace}
                    \\ &\qquad
                    - \int\!\!\!\dif{\mbf{s}_1}\dif{\mbf{s}_2}\;
                        \mcal{P}(\mbf{s}_1) \mcal{P}(\mbf{s}_2)
                        \DeltaT(\mbf{x}(\mbf{s}_1;\mbf{\nhat}))
                        \DeltaT(\mbf{x}(\mbf{s}_2;\mbf{\nhat}))
                \bigg]\nonumber \,,
    \end{align}
    where again $\dif{\mbf{s}}\equiv \dif{\spll}\dif{\sprp} \dif{\phi_s}$.

    Expectation values pass through integrals to apply only to the factors of
    $\DeltaT$, so that $\left\langle \Yd(\mbf{\nhat})\right\rangle$ can be
    expressed in terms of the correlation function $\xi\left(\norm{\mbf{x}_2 -
    \mbf{x}_1} \right)$ of the Gaussian $\DeltaT$ field:
    \begin{equation}
        \left\langle \Yd(\mbf{\nhat})\right\rangle
            \!=\! \frac{1}{2T_0^2 } \left[
                \xi(0) 
                \!-\!\! \int\!\!\!\dif{\mbf{s}_1}\dif{\mbf{s}_2}
                    \mcal{P}(\mbf{s}_1) \mcal{P}(\mbf{s}_2)
                    \xi(\Leq\norm{\mbf{s}_1 - \mbf{s}_2})
            \right],
    \end{equation}
    where we have used the fact that
    \begin{equation}
        \mbf{x}(\mbf{s}_1;\mbf{\nhat}) - \mbf{x}(\mbf{s}_2;\mbf{\nhat})
            = \Leq \, (\mbf{s}_1-\mbf{s}_2)\,.
    \end{equation}
    Unsurprisingly, $\langle\Yd\rangle$ is independent of direction, since we
    have assumed here that $\xi$ is statistically isotropic, at least on the
    small scales over which diffusion takes place during recombination:
    \begin{align} \label{eq:xiinrealspace}
        \xi(\norm{\mbf{x}_2 & - \mbf{x}_1})
            \equiv \left\langle \DeltaT(\mbf{x}_1) \DeltaT(\mbf{x}_2)\right\rangle
            \\
            &=
            \int\!\!\! \dif[3]{k_1} \dif[3]{k_2} \;
                \Exp{i(\mbf{k}_1\cdot\mbf{x}_1 - \mbf{k}_2\cdot\mbf{x}_2)}
                \langle A^T(\mbf{k}_1) A^{T*}(\mbf{k}_2) \rangle
            \\
            &=
            \int\!\!\! \dif[3]{k_1} \dif[3]{k_2} \;
                \Exp{i(\mbf{k}_1\cdot\mbf{x}_1 - \mbf{k}_2\cdot\mbf{x}_2)}
            \delta^{(3)}(\mbf{k}_1-\mbf{k}_2)
            P^{TT}(\norm{\mbf{k}_1})
            \\
            &=
            \int\!\!\! \dif[3]k \; \Exp{ i\mbf{k}\cdot(\mbf{x}_1-\mbf{x}_2) }
            P^{TT}(\norm{\mbf{k}}) 
    \end{align}
    The TT power spectrum $P^{TT}$ is the product of the initial power spectrum
    of the gauge-invariant curvature perturbations times the ``early''
    temperature transfer function squared
    \begin{equation}
        P^{TT}(\norm{\mbf{k}})
            = \frac{T_0^2}{4\pi}
              \frac{A_S(k_\star)}{k^3}
              \left(\frac{k}{k_\star}\right)^{n_s-1}
              ({\cal{T}}^e(k))^2
              \Exp{-2 (k/k_{2\rm{LS}})^2}\,.
    \end{equation}
    Here $k_\star=0.05 h^{-1}\rm{Mpc}$ is an arbitrary, but conventional, pivot
    scale.  The high-k cutoff $k_{2\rm{LS}}$, accounting for the damping up to
    second-last-scattering, can be written in terms of the conformal-time rate
    of change of the opacity \citep{Jungman+1996}, or more simply read directly
    from \citet[][eq. 7.139]{Baumann2022} 
    \begin{equation}
        k_{2\rm{LS}}^{-1} \simeq 8.8 \ \rm{Mpc}.
    \end{equation}

    \footnotetext{%
        Scattering transfers negligible energy to the photon in the rest frame
        of the electron, but more in a boosted frame.  In thermal equilibrium,
        the electron velocities due to thermal motions give a stationary
        photon-energy distribution.  So, to first order, we can neglect the
        effect of thermal velocities on the photon spectrum.  There is a
        second-order effect that causes a spectral distortion -- presumably
        ``toward'' a blackbody with temperature equal to the thermal electron
        temperature at the scattering point.  This will imprint an additional
        spectral distortion signal  due to the spatial variation of the
        temperature along each line of sight, but suppressed by the small
        fractional photon energy change at last scattering.%
    }

    We take the transfer function ${\cal{T}}^e(k)$ to be approximately the pure
    Sachs-Wolfe power spectrum, and, assuming that it changes slowly with time,
    evaluate it at the time of last scattering, rather than the time of
    second-last scattering \citep[we use][7.112]{Baumann2022}:
    \begin{align}
        \mcal{T}^e(k)
        &\simeq \left.{\mcal{T}}_{\rm SW}\right\vert_{\rm LS}(k) \label{eq:SWTransfer} \\
        &\simeq \frac{1}{5}
                \left[(1 + R(z_{\rm LS}))^{-1/4} \cos(k r_{s}(z_{\rm LS})) - 3R(z_{\rm LS})\right]
    \end{align}
    The label $e$ stands for early, i.e. during recombination, as opposed to the
    usual transfer function ${\cal{T}}^l(k)$, evaluated late, i.e. at redshift
    $z_0 = 100$.  Here
    \begin{equation}
        R(z) \equiv \frac{3}{4}\frac{\Omega_b}{\Omega_\gamma (1+z)} \simeq 0.6 \frac{\Omega_b h^2}{0.02}\frac{10^3}{1+z}
    \end{equation}
    and the sound horizon at emission \citep[see][Appendix C]{Baumann2022} is

    \begin{align}
         r_s (z_{\rm LS}&) = \int_\infty^{z_{\rm LS}} c_s \deriv[z]{\tau}\dif{z},
            \\
            &= \frac{2 c (1 \!+\! z_{\rm eq})}{3 H(z_{\rm eq})}
               \sqrt{\frac{6}{R_{\rm eq}}}
               \ln\left(\frac{\sqrt{1 + R(z)} + 
               \sqrt{R(z) + R_{\rm eq}}}{1 + \sqrt{R_{\rm eq}}}\right)
            \\
            &\simeq 144 \ \rm{Mpc}\,.
    \end{align}
    Thus,
    \begin{align} \label{eq:xiearlyofdeltar}
        &\xi^{ee}\left(r_{12} \right) \\
        &\ \simeq
            \frac{T_0^2 A_s(k_*)}{r_{12}} \!\!
            \int_{k_{\rm min}}^{\infty} \!\! \frac{\dif{k}}{k^2} 
                \left(\frac{k}{k_*}\right)^{n_s-1}
                \!\! \left(\mcal{T}^{e}_{\rm{SW}}(k) \Exp{-\left(\frac{k}{k_{2\rm{LS}}}\right)^2}\right)^2
                \sin(k r_{12}) \,.\nonumber
    \end{align}
    The result is independent of the low-$k$ cutoff for $k_{\rm min}\lesssim
    1\times10^{-2} \ \rm{Mpc}^{-1}$.

    Using the best current value \citep{Planck2018parameters} of
    $A_s\simeq2.1\times10^{-9}$ gives
    \begin{equation}
         \langle \Yd(\nhat)\rangle \simeq %
  \input{output/signal.txt}\label{output/signal.txt}\unskip%
 \,.
    \end{equation}
    Since the variance in the (dipole-subtracted) CMB temperature anisotropies,
    \begin{equation}
        \mathrm{var}(T)\Big\vert_{\mathrm{Planck}} = (15 \,\mu\mathrm{K})^2 \,,
    \end{equation}
    we might have thought that the maximum signal would be
    \begin{equation}
        \langle \Yd(\nhat)\rangle\vert_{\mathrm{expect}}
        = \frac{1}{2}\left(\frac{15\mu\mathrm{K}}{2.7 \mathrm{K}}\right)^2 \simeq 1.5\times 10^{-11}.
    \end{equation}
    However, the observed CMB fluctuations themselves are damped compared to
    their amplitude at second-last scattering.  Without that damping (i.e.
    eliminating the exponential term in \autoref{eq:xiearlyofdeltar}), the
    signal would have been
    \begin{equation}
         \langle \Yd(\nhat)\rangle\vert_{\mathrm{no-damping}} 
         = %
  \input{output/undamped_signal.txt}\label{output/undamped_signal.txt}\unskip%

    \end{equation}
    Damping of short range fluctuations is thus responsible for a
    factor-of-%
  \input{output/signal_ratio.txt}\label{output/signal_ratio.txt}\unskip%
 suppression in the signal from
    its maximum possible value.

% section calculating_the_signal (end)

\section{Correlation function of the signal and temperature fluctuations}
\label{sec:correlation_functions}

    Unlike the $y$-distortions that arise from the time-evolution of the
    background physics during recombination, the $y$-distortions due to local
    blackbody averaging are anisotropic.  We are therefore interested in
    calculating the angular correlation function of $\Yd(\nhat)$ with the
    temperature fluctuations themselves $\DeltaT(\nhat')$.  The particular form
    of this cross-correlation is distinctive to the diffusion distortion signal,
    and can be used to disentangle diffusion distortion from foregrounds.

    Since the fluctuations are nearly Gaussian, the expected correlation of
    $\Yd(\nhat)$ with $\DeltaT(\nhat')$ is a three-point function, and nearly
    vanishes.  However, the correlation of $\Yd(\nhat)$ with
    $(\DeltaT(\nhat'))^2$ does not:

    \begin{align}
        & \frac{1}{T_0^2}\Yd(\nhat_1)(\DeltaT(\nhat_2))^2
            \\
            &\quad =
             \frac{1}{2T_0^4}
                \Bigg[
                    \ \phantom{+} \int\!\!\!\dif{\mbf{s}_1}\;
                        \mcal{P}(\mbf{s}_1) 
                        \left(\DeltaT(\mbf{x}(\nhat_1,\mbf{s}_1)\right)^2
                 \nonumber\\ &\qquad\qquad\
                    - \int\!\!\!\dif{\mbf{s}_1}\; \mcal{P}(\mbf{s}_1) \dif{\mbf{s}'_1} \mcal{P}(\mbf{s}'_1)
                        \DeltaT(\mbf{x}(\nhat_1,\mbf{s}_1)
                        \DeltaT(\mbf{x}(\nhat_1,\mbf{s}'_1)
                \Bigg] \nonumber
                \\ &\qquad\quad
                \times \int\!\!\!\dif{\mbf{s}_2} \; \mcal{P}(\mbf{s}_2) \dif{\mbf{s}'_2} \mcal{P}(\mbf{s}'_2)
                    \DeltaT(\mbf{x}(\nhat_2,\mbf{s}_2)
                    \DeltaT(\mbf{x}(\nhat_2,\mbf{s}'_2) \nonumber
    \end{align}
    so
    \begin{align}
    & C^{\Yd(\DeltaT)^2}(\mbf{\nhat}_1,\mbf{\nhat}_2) \nonumber
        \\
        &\quad \equiv
            \Big\langle \frac{1}{T_0^2}\Yd(\nhat_1)(\DeltaT(\nhat_2))^2 \Big\rangle
            - \frac{1}{T_0^2} \Big\langle\Yd(\nhat_1)\Big\rangle \Big\langle(\DeltaT(\nhat_2))^2 \Big\rangle 
        \\
        &\quad =
            \phantom{+} \frac{1}{T_0^4}
            \int\!\!\!\dif{\mbf{s}_1} \mcal{P}(\mbf{s}_1) \dif{\mbf{s}_2} \mcal{P}(\mbf{s}_2) \dif{\mbf{s}'_2} \mcal{P}(\mbf{s}'_2)
            \\ &\qquad\qquad
            \xi^{el}\left(\norm{ \mbf{x}(\nhat_1,\mbf{s}_1) - \mbf{x}(\nhat_2,\mbf{s}_2) }\right)
            \xi^{el}\left(\norm{ \mbf{x}(\nhat_1,\mbf{s}_1) - \mbf{x}(\nhat_2,\mbf{s}'_2) }\right) \nonumber
            \\ &\qquad
            - \frac{1}{T_0^4}
              \left(
                \int\!\!\! \dif{\mbf{s}_1} \mcal{P}(\mbf{s}_1) \dif{\mbf{s}_2} \mcal{P}(\mbf{s}_2)
                    \xi^{el}\left(\norm{ \mbf{x}(\nhat_1,\mbf{s}_1) - \mbf{x}(\nhat_2,\mbf{s}_2) }\right)
              \right)^2 \,. \nonumber
    \end{align}
    (See \appref{app:calculating_the_correlation} for a full derivation.)

    Given a statistically isotropic universe, $C^{\Yd(\Delta
    T)^2}(\mbf{\nhat}_1,\mbf{\nhat}_2)$ depends only on
    $\nhat_1\cdot\nhat_2\equiv\cos\theta$.  Here
    \begin{align}\label{eqn:xiearlylateofdeltar}
        \xi^{el}\left(r_{12} \right) 
        &\simeq
            \frac{T_0^2 A_s(k_*)}{r_{12}}
               \int_{k_{\rm min}}^{\infty} \!\! \frac{\dif{k}}{k^2} 
                    \left(\frac{k}{k_*}\right)^{n_s-1}
                    \!\!\!\!\!\!\!\! \mcal{T}^{e}(k)
                    \mcal{T}^{l}(k) \\
            &\qquad \qquad \qquad \qquad \qquad \times 
                \Exp{-2 (k/k_{2LS})^2} \sin(k r_{12})   \,,\nonumber
    \end{align}
    with ${\cal{T}}^l(k) $ obtained by evaluating ${\cal{T}}_{SW}(k)$ from
    \eqref{eq:SWTransfer} at $z_0$.

\begin{figure}
        \centering
        \includegraphics[width=1\linewidth]{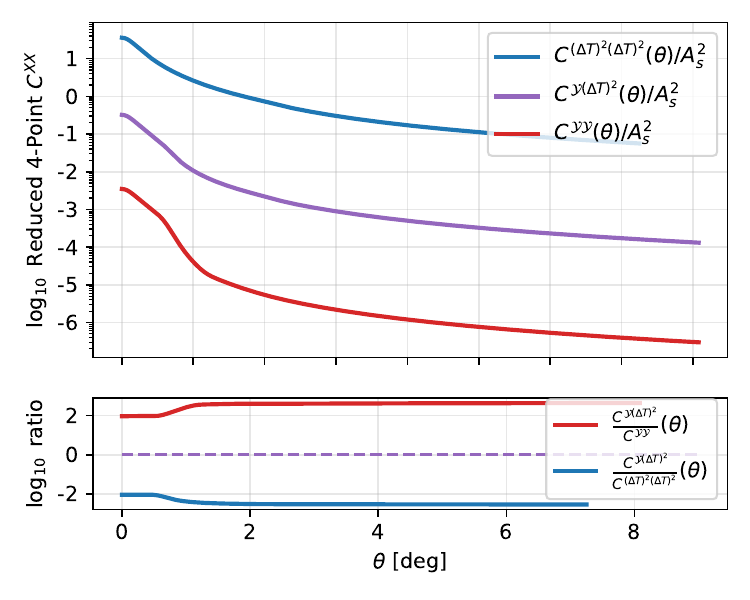}
        \caption{%
            \textbf{Upper panel}: Reduced dimensionless angular correlation
            functions of $Y$, $C^{\mcal{YY}}(\theta)/A_s^2$ (red), between $Y$
            and $(\DeltaT)^2$, $C^{\Yd(\DeltaT)^2}(\theta)/A_s^2$ (purple), and
            between the Sachs-Wolfe contributions to $(\DeltaT)^2$,
            $C^{(\DeltaT)^2(\DeltaT)^2}(\theta)/A_s^2$ (blue), where $\theta$ is
            the angle between sky locations of the two quantities being
            correlated.  \textbf{Lower panel}: the ratio of $C^{\Yd(\DeltaT)^2}$
            to $C^{(\DeltaT)^2(\DeltaT)^2}$ (blue) and $C^{\Yd(\DeltaT)^2}$ to
            $C^{\Yd\Yd}$ (red).
        }
        \label{fig:y-tsqcorrelation}
        \script{correlation/plot_correlation.py}
    \end{figure}

    In the upper panel of \autoref{fig:y-tsqcorrelation}, we plot
    $C^{\Yd(\DeltaT)^2}(\theta)$ in purple and, for comparison (in blue), the
    pure-SW $(\DeltaT)^2-(\DeltaT)^2$ auto-correlation function
    \begin{align}
        C&^{(\DeltaT)^2 (\DeltaT)^2}(\mbf{\nhat}_1\cdot\mbf{\nhat}_2) \nonumber
        \\
        &\equiv
            \frac{1}{T_0^4}
            \left(\!
                \Big\langle \!(\DeltaT(\nhat_1))^2(\DeltaT(\nhat_2))^2 \Big\rangle
                \!-\! \Big\langle\!(\DeltaT(\nhat_1))^2\Big\rangle \Big\langle\!(\DeltaT(\nhat_2))^2 \Big\rangle
            \!\right)
        \\
        &=
            \phantom{+}\frac{1}{T_0^4}
            \int\!\!\! \dif{\mbf{s}_1} \; \mcal{P}(\mbf{s}_1) \dif{\mbf{s}_2} \mcal{P}(\mbf{s}_2) \dif{\mbf{s}'_2} \mcal{P}(\mbf{s}'_2)
            \\
            &\qquad\qquad
                \xi^{ll}\left(\norm{ \mbf{x}(\nhat_1,\mbf{s}_1) - \mbf{x}(\nhat_2,\mbf{s}_2) }\right)
                \xi^{ll}\left(\norm{ \mbf{x}(\nhat_1,\mbf{s}_1) - \mbf{x}(\nhat_2,\mbf{s}'_2) }\right) \nonumber 
            \\ &\qquad
                    - \frac{1}{T_0^4} \left( \int\!\!\! \dif{\mbf{s}_1} \; \mcal{P}(\mbf{s}_1) \dif{\mbf{s}_2} \mcal{P}(\mbf{s}_2)
                        \xi^{ll}\left(\norm{\mbf{x}(\nhat_1,\mbf{s}_1) - \mbf{x}(\nhat_2,\mbf{s}_2) }\right)
                        \right)^2
                    \nonumber\,.
    \end{align}
    $C^{(\DeltaT)^2(\DeltaT)^2}$ is also a function only of
    $\cos\theta=\nhat_1\cdot\nhat_2$.  
    We also plot (in red) $C^{\Yd\Yd}$.  The small-scales wiggles in the power
    spectra are numerical artifacts of the discrete Legendre transform, and are
    immaterial to our conclusions.  The correlation functions in
    \autoref{fig:y-tsqcorrelation} are smooth, as expected. 

    \begin{figure}
        \includegraphics[width=1\linewidth]{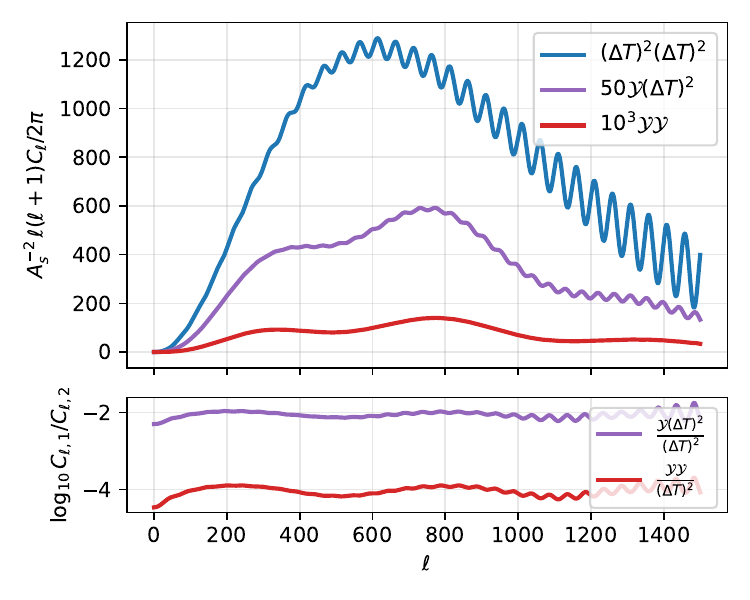}
        \caption{% 
            \textbf{Upper panel}: Legendre expansion coefficients (cf.
            \eqref{eq:LegendreCoefficents}) of the correlation functions
            $C^{\mcal{YY}}(\theta)/A_s^2$ (red) and
            $C^{\Yd-(\DeltaT)^2}(\theta)/A_s^2$ (purple), compared to those of
            $C^{(\DeltaT)^2-(\DeltaT)^2}(\theta)/A_s^2$ (blue), accounting
            solely for Sachs-Wolfe fluctuations. The small-scale wiggles are
            numerical artifacts of the discrete Legendre transformation and are
            not material to our conclusions. \textbf{Lower panel}: the ratio of
            the expansion coefficients to those of $(\DeltaT)^2(\DeltaT)^2$.%
        }
        \label{fig:y-tsqangularPS}
        \script{cls/plot_cls.py}
    \end{figure}
    
    In the lower panel of the figure, we plot (in blue) the ratio of
    $C^{\Yd(\DeltaT)^2}$ to $C^{(\DeltaT)^2(\Delta T)^2}$, as well as the ratio
    of $C^{\Yd(\DeltaT)^2}$ to $C^{\Yd\Yd}$ (in red).  In
    \autoref{fig:y-tsqangularPS}, we plot the corresponding Legendre expansion
    coefficients $C_\ell$, obtained by fitting $C_\ell$ to
    \begin{equation} \label{eq:LegendreCoefficents}
        C^{AB}(\theta) = \sum\limits_{\ell=\ell_{\rm min}}^{\ell_{\rm max}} 
            \frac{\ell(\ell+1)}{4\pi}C^{AB}_\ell P_\ell(\cos\theta)
    \end{equation}
    for $C^{AB}_\ell$, with $A=\Yd,(\DeltaT)^2$ and $B=(\DeltaT)^2$.
    
    Over substantial ranges of angle, $C^{\Yd(\DeltaT)^2}>
    10^{-3}C^{(\DeltaT)^2(\DeltaT)^2}$.  Similarly, over a sizable range of
    $\ell$, $C_\ell^{\Yd(\DeltaT)^2} > 10^{-3}C_\ell^{(\Delta T)^2(\DeltaT)^2}$.

\section{Detectability} \label{sec:detectability}

    The $\mcal Y$ power spectrum is well below the detection capability of
    envisioned experiments, and is small compared to $\mathcal Y$ from
    low-redshift contributions.  But the ${\mcal Y}(\DeltaT)^2$ cross power
    spectrum is more promising, because it has both a significantly larger
    amplitude and a distinctive shape, including slight imprints of acoustic
    oscillations, which can distinguish it from other contributions.

    We estimate the detectability of $C_\ell^\Yd(\DeltaT)^2$ by assuming that
    both $\Yd$ and $(\DeltaT)^2$ are Gaussian random fields on the sky.  While
    this is not precisely the case, it is a large simplification and should be
    sufficient for a reasonable signal-to-noise estimate.  If the CMB blackbody
    temperature in a given sky pixel can be measured with an uncertainty of
    $\delta_T$, then the uncertainty in $(\DeltaT)^2$ is $2\DeltaT\delta_T +
    \delta_T^2$.  Any map with a chance to detect the signal described here will
    have $\delta_T \ll \DeltaT$ so the second term can be dropped.  Since
    $\DeltaT$ is itself normally distributed, it can be replaced by its rms
    value, so the uncertainty in $(\DeltaT)^2$ is $\delta_{T^2} \simeq
    2\DeltaT_{\rm rms}\delta_T$.  For a map of $\Yd$, each pixel corresponds to
    a particular intensity distortion at each frequency, given by the usual
    thermal Sunyaev-Zeldovich formula.  Taking 90 GHz as a typical frequency for
    ground-based experiments, an uncertainty in temperature $\delta_T$ gives
    $\delta_Y = 1.6 \delta_T / T_0$ as the uncertainty in $\Yd$.

    Given these pixel errors in two (approximately) Gaussian random fields, the
    uncertainty in measuring each $\ell$ mode of the angular cross-power
    spectrum $C_\ell^{\Yd(\DeltaT)^2}$ is analogous to determining $C_\ell^{TE}$
    from maps of microwave background temperature and E-mode polarization.  The
    variance, including both pixel noise and cosmic variance, is given by
    \citealt[Eq.~(3.26)]{Kamionkowski+1997}, as
    %
    % \vskip 20pt
    \begin{widetext}
        \begin{equation}
            \Xi_{\Yd(\DeltaT)^2,\Yd(\DeltaT)^2}
                = \frac{1}{2\ell+1} 
                  \left[ 
                      \left(C_\ell^{\Yd(\DeltaT)^2}\right)^2
                    + \left(C_\ell^{(\DeltaT)^2(\DeltaT)^2} + w^{-1}_{(\DeltaT)^2} \left(W_\ell^b\right)^{-2}\right)
                        \left(C_\ell^{\Yd\Yd} + w^{-1}_{\Yd} \left(W_\ell^b\right)^{-2}\right)
                  \right]\,.
        \end{equation}
        Here $w_{X}^{-1} \equiv 4\pi(\delta_{X}^2)/N_{\rm pix}$ is the inverse
        statistical weight per unit solid angle on the sky for a map of some
        quantity with $N_{\rm pix}$ pixels and a pixel variance of $\delta^2$,
        and $W_\ell^b = \exp(-\ell^2\sigma_b^2/2)$ is the experiment beam
        profile in harmonic space, which is generally well approximated as a
        Gaussian with beam with $\sigma_b = \theta_{\rm fwhm} / \sqrt{8\ln 2} =
        1.2\times 10^{-4}(\theta_{\rm fwhm}/ 1')$.  The signal-to-noise with
        which a given $\ell$ mode can be measured is then just
        \begin{equation}
            \frac{\rm S}{\rm N}\left(C_\ell^{\Yd (\DeltaT)^2}\right)
                = \sqrt{(2\ell + 1)f_{\rm sky}}
                  \left[
                    1
                    + \left(
                        \frac{C_\ell^{(\DeltaT)^2(\DeltaT)^2}}{C_\ell^{\Yd (\DeltaT)^2}}
                        + \frac{w_{T^2}^{-1} \left(W_\ell^b\right)^{-2}} {C_\ell^{\Yd (\DeltaT)^2}}
                      \right)
                      \left(
                        \frac{C_\ell^{\Yd \Yd}}{C_\ell^{\Yd (\DeltaT)^2}}  
                        + \frac{w_{\Yd}^{-1} \left(W_\ell^b\right)^{-2}} {C_\ell^{\Yd (\DeltaT)^2}}
                      \right)
                  \right]^{-1/2}\,,
        \end{equation}
        \label{eq:cl_sn}
    \end{widetext}
    where a factor of $f_{\rm sky}$, giving the sky fraction
    covered by a map, has been included.  To the extent the fields $\Yd$ and
    $(\DeltaT)^2$ are Gaussian, each mode is statistically independent, and the
    total signal-to-noise can be obtained by summing over all $\ell$ values
    probed by a given map.

    As an example of the current experimental state of the art, we consider the
    ACT experiment \citep{Coulton+2023}; maps of similar statistical weight and
    sky coverage have also been made by the SPT experiment \citep{SPTymap}.
    Current ACT maps of the blackbody temperature component have uncertainties
    ranging from 5 $\mu$K to 14 $\mu$K per square arcminute; we take $\delta_T =
    10$ $\mu$K as a (conservative) mean temperature uncertainty.  The final ACT
    data set provides maps with $f_{\rm sky} = 0.33$, so for $(1')^2$ pixels,
    $N_{\rm pix} = 4.8\times 10^7$.  The statistical weight factors are
    therefore 
    % $4\pi(10\mu{K})^2/(4.8\times 10^7)$
    $w^{-1}_T = 2.6\times 10^{-5}$ $\mu{\rm K}^2$ and
    % $4\pi(2(Delta T)_{rms}\delta_T)^2/N_{pix}  = 4\pi(2~15\mu{K}
    % 10\mu{K})^2/(4.8\times 10^7) $
    $w^{-1}_{(\DeltaT)^2}=0.02$ $\mu{\rm K}^4$ (using $(\Delta T)_{\rm rms} =
    15\mu{\rm K}$).  Dividing by factors of $T_0$ give the dimensionless values
    $w^{-1}_T =3.6\times 10^{-18}$ and $w^{-1}_{(\DeltaT)^2}= 3.8\times
    10^{-28}$.  Meanwhile $w^{-1}_{\cal Y} = (1.6)^2 w^{-1}_T$.  ACT considers
    measured $\ell$ values below about 500 to be potentially unreliable, so we
    only use multipole moments  
    $\ell > 500$.

    For these parameters, the terms in \hyperref[eq:cl_sn]{eq. 54} containing
    $w_{(\DeltaT)^2}^{-1}$ and $C_\ell^{\Yd \Yd}$ can be neglected, along with
    the $1$, so the expression simplifies to 
    \begin{equation}
        \frac{\rm S}{\rm N}\left(C_\ell^{\Yd (\DeltaT)^2}\right)
            = \left(\frac{(2\ell + 1)f_{\rm sky}}{C_\ell^{(\DeltaT)^2(\DeltaT)^2}w_{\Yd}^{-1}}\right)^{1/2} 
            C_\ell^{\Yd (\DeltaT)^2}W_\ell^b.
    \end{equation}
    Summing from $\ell=500$ to $\ell=1500$ gives an expected total S/N of
    $%
  \input{output/sn_act.txt}\label{output/sn_act.txt}\unskip%
$ for ACT.  So the diffusion spectral
    distortion should be detectable at current experiment sensitivities.  SPT
    has released \citep{SPTymap} a y-distortion map with $1.25'$ angular
    resolution, and approximately comparable quality to what we have assumed for
    ACT (although with a smaller $f_{\rm sky}$).  The planned CMB-S4 experiment
    might improve on this by a factor of 10, principally by reducing $\delta_T$
    to $\simeq 1\mu$K.  This would allow robust detection of the acoustic
    oscillations in $C_\ell^{\Yd (\DeltaT)^2}$, and might therefore allow ${\cal
    Y}$ to be used as a cosmological probe analogous to polarization.

    In forecasting the S/N, we have made several assumptions.  Primarily, we
    have assumed that ${\cal Y}$ can be detected with uncertainty comparable to
    that in $\Delta T$.  That requires robust foreground subtraction, since
    $\Yd$ is large in the location of galaxy clusters, and the $\Yd$ map will be
    dominated by distortions from foreground haloes.  Since we are correlating
    with $(\DeltaT)^2$, modeling this signal due to foreground $\Yd$
    contributions will likely be the primary challenge in extracting the
    recombination-era signal.  We have also worked in the approximation that
    ${\cal Y}$ and $(\Delta T)^2$ are Gaussian, which they are not, and a more
    careful statistical treatment is merited.  Finally, we have included only
    the Sachs-Wolfe contribution to the transfer function, and used an analytic
    approximation.

% section: detectability (end)

\section{Discussion and Conclusions} \label{sec:conclusion}

    In this paper we provide a real-space description of the $y$-type spectral
    distortion of the CMB that arises from the scattering of the photons into
    our line of sight \citep{Zeldovich+1972}.  Each photon's scattering history
    is different, sampling both radially along and transverse to the line of
    sight through the recombination era.  Because the photons' final scatterings
    make nearly no change to the photon energy, the resulting distribution of
    photon energies that we observe is a mixture of blackbody distributions of
    different temperatures, representing the inhomogeneity of the temperature in
    the region from which the photons originated.

    This ``diffusion spectral distortion'' is, like the CMB intensity
    (temperature) and polarization, a probe of the acoustic modes responsible
    for the inhomogeneities in the universe during the epoch of recombination.
    Like the E-mode polarization and the temperature, in standard $\Lambda$ Cold
    Dark Matter cosmology, the diffusion spectral distortion signal is partly,
    but not perfectly, correlated with these other signals.  For example, while
    the E-mode fluctuations probe the local quadrupole of the temperature
    distribution, the diffusion spectral distortion is also sensitive to its
    dipole.

    Diffusion spectral distortion offers another independent probe of the
    physics at the end of recombination.  Inherently, this implies the
    opportunity to reduce cosmic variance on existing measurements of quantities
    probed by the temperature and polarization.  Additionally, because it is
    sensitive to the variation in the photon temperature along the
    line-of-sight, diffusion spectral distortion could potentially be the basis
    of a new Alcock-Paczynski test \citep{AlcockPacynski1979} at the epoch of
    recombination.  It could also allow us to test statistical isotropy at the
    epoch of recombination, complementing the large scale tests traditionally
    done with temperature that have yielded anomalous results
    \citep{Schwarz+2015, Planck2018stats, Abdalla+2022}.  Diffusion
    spectral distortion is therefore both a consistency check for the standard
    cosmological model and sensitive to new physics in a way that is
    complementary to other signals.

    Polarization is also generated during recombination, and usually described
    as a measure of the local quadrupole at the last scattering.  The
    polarization spectrum and its distortion away from blackbody are therefore
    due to a different weighted sum over blackbodies, with the potential for yet
    another signal complementary to the one we have described.  This is another
    step towards a possible tomographic probe \citep{YadavWandelt2005} of the
    universe through recombination.

    Current and upcoming experiments will not be sufficiently sensitive to the
    $y$-type distortion to measure the diffusion spectral distortion signal
    directly.  In part, the challenge is insufficient signal-to-noise.  However,
    a greater challenge is likely to be separation of the component of the
    $y$-distortion due to recombination-era diffusion from other causes of
    spectral distortion in the presence of foregrounds with spectra that are
    imperfectly known and that are also anisotropic \citep{Abitbol+2017,
    Hart+2020, RottiChluba2021}.  These foregrounds include our Milky Way, and
    both galaxies and galaxy clusters at all redshifts.  The $\Yd$
    auto-correlation function, like the $y$-type distortion, will not be
    detectable by current or upcoming experiments.  The amplitude of the
    correlation signal is far smaller than for other sources of $y$
    auto-correlation at low redshifts, and so will be difficult to separate from
    foregrounds.
    
    More promising than the $\Yd$ signal itself or its auto-correlation function
    is the cross-correlation $C_\ell^{\Yd (\DeltaT)^2}$ between the
    $y$-distortion and the square of the temperature fluctuations.  
    (The $\Yd$-$T$ correlation will be zero if the primordial photon
    perturbations are a Gaussian field.) This is calculated in
    \autoref{sec:correlation_functions}.  Like the temperature fluctuations, the
    diffusion spectral distortion, and hence their cross-correlation function,
    contains acoustic features.  The combination of the specific spectral shape
    and the correlation with the primordial temperature fluctuations should
    facilitate the separation of this correlation function from foreground
    signals.  Galactic foreground confusion is uncorrelated with primordial
    temperature anisotropies (though of course it would be correlated with any
    residual unsubtracted Galactic temperature foreground).  Extragalactic
    foregrounds are concentrated at galaxy clusters (which are detected at high
    significance) and at galaxies.  Fortunately, the clusters can be masked and
    are relatively sparse on the sky.  Galaxies are far more numerous and
    non-sparsely distributed; however, the amplitude of the galaxy confusion
    limit $y$-distortion is also small and would have a different correlation
    function with the temperature fluctuations.

    The three-point correlation function between $\Yd(\nhat_1)$,
    $\DeltaT(\nhat_2)$, and $\DeltaT(\nhat_3)$ at three different points may
    also be detectably large.  Considering different configurations of this
    three-point function will give further handles on separating the diffusion
    distortion from various foreground contributions \citep{Coulton+2018}.
    Calculation of this signal is more complicated than the two-point function
    and will be considered elsewhere.

    In \autoref{sec:detectability} we determined that multi-frequency maps over
    a third of the sky with the $10\mu$K--arcmin sensitivity attained by the
    final ACT dataset \citep{Coulton+2023} have sufficient sensitivity to
    detect this signal, and SPT has achieved similar reach \citep{SPTymap}.
    Sky maps with sensitivities approaching 1 $\mu$K--arcmin are anticipated in
    the next decade \citep{Abitbol+2017cmbs4}.  Sufficient frequency coverage is
    required to separate the $y$-distortion from the primary blackbody and other
    spectrum components.

    Our signal and sensitivity calculations involved several simplifying
    assumptions.  More rigorous calculations of the $y$-distortion and its
    correlation with the temperature and polarization anisotropies during
    diffusion damping are warranted.  Ultimately, a complete calculation of the
    statistics of spectral distortions arising from physical processes around
    last scattering may reveal additional probes of the cosmological model,
    providing substantial consistency checks or additional handles on
    non-standard physics.

% section conclusions (end)

% -------------------------------------------------------------------

\section*{Acknowledgements} \label{sec:acknowledgements}

    N.S. acknowledges support from the Natural Sciences and Engineering Research
    Council of Canada (NSERC) - Canadian Graduate Scholarships Doctorate Program
    [funding reference number 547219 - 2020].  N.S. received partial support
    from NSERC (funding reference number RGPIN-2020-04712) and from an Ontario
    Early Researcher Award (ER16-12-061; PI Bovy).  N.S. would also like thank
    Prof. Jeremy Webb for providing computational resources.
    G.D.S. was partially supported by DOE grant DESC0009946.  G.D.S. thanks
    Imperial College London for hospitality while some of this work was
    completed.

    The data availability statement is modified from one provided to
    \textsc{showyourwork} by Mathieu Renzo.

    \vspace{20pt}
    \paragraph*{Software (alphabetical)}

        \package{asdf} \citep{Greenfield+2015}, %
        \package{astropy} \citep{Astropy+2013, Astropy+2018, Astropy+2022}, %
        \package{CLASS} \citep{CLASS1, CLASS2} %
        \package{Cosmology-API} \citep{cosmology-api} %
        \package{interpolated-coordinates} \citep{interpolated-coordinates} %
        \package{Matplotlib} \citep{Hunter2007}, %
        \package{NumPy} \citep{Harris+2020}, %
        \package{SciPy} \citep{Scipy2020}, %
        \package{ShowYourWork} \citep{Luger+2021} %

% section acknowledgements (end)

% -------------------------------------------------------------------

\clearpage
\section*{Data Availability} \label{sec:data_availability}

    This study was carried out using the reproducibility software
    \href{https://github.com/showyourwork/showyourwork}{\showyourwork}
    \citep{Luger+2021}, which uses continuous integration to programmatically
    download the data, perform the analyses, create the figures, and compile the
    manuscript.  Each figure caption contains two links: one to the dataset used
    in the corresponding figure, and the other to the script used to make the
    figure.  The datasets are stored at \url{https://zenodo.org/record/8400583}.
    The git repository associated with this study is publicly available at
    \href{https://github.com/nstarman/Temperature-Diffusion-Spectral-Distortion-Paper/}{\UrlFont{Temperature-Diffusion-Spectral-Distortion-Paper}}.

% section data_availability (end)

%%%%%%%%%%%%%%%%%%%%%%%%%%%%%%%%%%%%%%%%%%%%%%%%%%%%%%%%%%%%%%%%%%%%%%%%%%%%%%%
% REFERENCES

\bibliographystyle{mnras}
\bibliography{starkman_etal_2023_diffusion_distortion}

%%%%%%%%%%%%%%%%%%%%%%%%%%%%%%%%%%%%%%%%%%%%%%%%%%%%%%%%%%%%%%%%%%%%%%%%%%%%%%%
% Appendices

\appendix
\onecolumn

\section[Coordinate Systems]{Coordinate Systems}\label{app:coordinate_systems}

    We use the standard definition of the scale factor: $a(z=\infty) = 0$,
    $a(z=0)=1$.

    For analytic simplicity, we consider a two component universe containing
    only matter and radiation, whose energy densities were equal when the value
    of the scale factor was $\aeq$, or equivalently at redshift $z_{\rm eq}$.
    This is an excellent approximation during the epoch of recombination and
    last scattering, when all the complex physics of this problem takes place.
    Of course it is a poor approximation thereafter, but that is easily
    accounted for by placing a reference observer at a redshift much less than
    that of last scattering, but much greater than that of cosmological constant
    dominance.  We are therefore able to take $z_{\rm eq}$  to be its inferred
    value from observations; in particular $z_{\rm eq}=(\Omega_m/\Omega_r - 1) =
    3404$ from \cite{Planck2018parameters}.

    The comoving distance along a photon's path from a point a with scale factor
    $a_i$ to a point with scale factor $a_f>a_i$ is
    \begin{align}
        r(a_i, a_f) &= \frac{\Leq}{\sqrt{8 \aeq} } \int_{a_i}^{a_f}\frac{\dif{a'}}{\sqrt{\aeq+a'}}
        =\Leq \left.\sqrt{\frac{1+a/\aeq}{2}}\right\vert_{a_i}^{a_f} \,,
        \label{eq:path_length}
    \end{align}
    where 
    \begin{equation}
        \Leq\equiv \frac{c}{H_0}\sqrt{\frac{8\, \aeq}{\OmO}}
    \end{equation}

    It proves convenient, for calculating $\mcal{P}(\spll,\sprp,\phi)$
    \autoref{eq:Pspllsprpphi}, to use dimensionless comoving distances that are
    anchored, time-oriented, and of a convenient magnitude during recombination.
    We define
    \begin{equation} \label{eq:rho_of_z}
        \rho(z) \equiv \sqrt{\frac{1}{2} \left(1+\frac{1+z_{\rm eq}}{1+z} \right)}\,.
    \end{equation}
    Useful special cases are: $\rho_{\rm eq}=1$, and $\rho(z=\infty)
    \equiv1/\sqrt{2}$.  Because we are using a matter-plus-radiation
    approximation for  the evolution of the scale factor, we cannot use our
    definition of $\rho$ for late times, for example today ($z=0$).  We take
    $\rho_o$ to be the largest allowed value of $\rho$, taken at $z=z_o=100$.
    Now,
    \begin{equation}
        r(z_i,z_f)
         =\Leq\;(\rho(z_f)-\rho(z_i))
    \end{equation}

% section coordinate_systems (end)

\section{Visibility Function } \label{app:visibility_function}

    The epoch of recombination was not instantaneous, and the universe did not
    become transparent instantaneously, so CMB photons did not propagate
    unimpeded from some fixed epoch.  The visibility function is the probability
    per-unit-distance that a photon observed at position $x_2$ last interacted
    at position $x_1$.  \citet[eq. 1-3]{Abramo+2010} gives a very good
    explanation of the visibility function, phrased in terms of the conformal
    time $\eta$, with $a d\eta\equiv dt$.  We list the important terms, noting
    small changes to notation and that we express quantities as functions of
    redshift not conformal time:
    \begin{itemize}[leftmargin=2\parindent]
        \item $\bar{\mu}(z_1, z_2)$, the optical depth for Thompson scattering;
        \item
            $\PgamBar(z_1, z_2) = \Exp{-\bar{\mu}(z_1, z_2)}$, the total
            visibility;
        \item
            $\GgamBar(z_1, z_2) = \deriv[\eta_1]{z_1} \deriv[z_1]{}
            \PgamBar(z_1, z_2)$, the visibility function, which is the
            likelihood of a photon Thompson scattering between redshifts $z_1$
            and $z_2$.
    \end{itemize}
    The overbar in $\bar{\mu}$, $\PgamBar$, and $\GgamBar$ indicates that they
    are functions only of $z$ and not position, as they refer to the unperturbed
    ``background cosmology".

    For numerical and analytic purposes, we want to ``split" our background
    quantities, i.e. we rewrite $\PgamBar(z_1,z_2)$ (where $z_2<z_1)$ in terms
    of CLASS's $\PgamBarCL(z)$ which is anchored at the observer $z_O$:
    \begin{align}
        \PgamBarCL(z) &= \PgamBar(z,z_O) = \Exp{-\mu(z,z_O)} & \nonumber
    \intertext{\quad For this we write\newline}
        \PgamBar(z_1,z_2)
        &= \frac{\PgamBar(z_1,z_O)}{\PgamBar(z_2,z_O)}
        = \frac{\PgamBarCL(z_1)}{\PgamBarCL(z_2)} &  \label{eq:PgamBarCl}\,.
    \end{align}
    
    Similarly, for $\GgamBar$:

    \begin{align}
        \GgamBarCL(z) &= \GgamBar(z,z_O)\,, &
    \intertext{\quad and so\newline}
        \GgamBar(z_1,z_2) 
        &= \deriv[\eta_1]{z_1} \deriv*[z_1]{\Exp{-\mu(z_1,z_2)}}
        = \deriv[\eta_1]{z_1} \deriv*[z_1]{\frac{\Exp{-\mu(z_1,z_O)}}{\Exp{-\mu(z_2,z_O)}}}
        = \frac{1}{\Exp{-\mu(z_2,z_O)}}
        \deriv[\eta_1]{z_1} \deriv*[z_1]{\Exp{-\mu(z_1,z_O)}}
        = \frac{\GgamBarCL(z_1)}{\PgamBarCL(z_2)}\,. & \label{eq:GgamBarCl}
    \end{align}

% section visibility_function (end)

\section[Calculating P]{Calculating and Sampling $\mcal{P}(\spll, \sprp,
\phi)$}\label{app:Pspllsprp}

    For calculating and sampling $\mcal{P}$, it is convenient to perform a
    change of coordinates.  In \autoref{eq:s_definition} we define $\spll,
    \sprp$, and in \autoref{eq:rho_of_z} we define $\rho$.  We define
    \begin{equation}
        \GgamBar(\rho_2,\rho_1) \equiv \GgamBar(z_2,z_1)\,,
    \end{equation}
    normalized such that $\Leq \int\!\dif{\rho_2} \; \GgamBar(\rho_2,\rho_1)=1$.
    
    Rewriting \autoref{eq:Pspllsprpphi} in terms of $\rho_i$ and the
    CLASS-defined functions,
    \begin{align} \label{eq:Pspllsprpphi_app}
        \mcal{P}(\spll,\sprp,\phi)
            &= \frac{3\Leq^2}{16\pi\PgamBarCL(\rho_O)}
                \int\!\!\! \dif{\rho_1} \; \GgamBarCL(\rho_1) \frac
                {\GgamBarCL\left(\rho_1\! - \! \sqrt{(\rho_1+\spll-\rho_O)^2+ \sprp^2}\right)}{\PgamBarCL(\rho_1)}
                \frac{(2(\rho_1+\spll-\rho_O)^2 + \sprp^2)\sprp}{\left((\rho_1+\spll-\rho_O)^2+\sprp^2\right)^2} \,,
    \end{align} 
    which we evaluate as follows.  For a given $\spll, \sprp$ we perform a cubic
    spline in $\rho_1$ of 
    \begin{equation}\label{eq:spline_gPg}
        f(\rho_1;\spll-\rho_O,\sprp) \equiv 
        \frac{\GgamBarCL(\rho_1)} {\PgamBarCL(\rho_1)}
                {\GgamBarCL\left(\rho_1\! - \! \sqrt{(\rho_1+\spll-\rho_O)^2+ \sprp^2}\right)} \,.
    \end{equation}
    Between knots of the spline ($\rho_1\in[\rho_1^{(i)},\rho_1^{(i)}]$), we
    must do the integral
    \begin{equation}\label{eq:spline_integral_P}
        \sum_{j=0}^3 f_j^{(i)}(\spll-\rho_O,\sprp) 
         \int_{\rho_1^{(i)}}^{\rho_1^{(i+1)}}\!\!\!\!\! \dif{\rho_1} \;
         \rho_1^j~
         \frac{(2(\rho_1+\spll-\rho_O)^2 +
                \sprp^2)\sprp}{\left(( \rho_1+\spll-\rho_O)^2+\sprp^2\right)^2} \,.
    \end{equation}
    The integrals with different $\rho_1^j$ can all be done analytically:
    \begin{align}
       \bullet \int\!\dif{\rho_1}\, \rho_1^0~ \frac{(2(\rho_1+\spll-\rho_O)^2 + \sprp^2)\sprp}{\left(( \rho_1+\spll-\rho_O)^2+\sprp^2\right)^2}
            &= - \frac{\sprp(\rho_1+\spll-\rho_O)}
                      {2 \left((\rho_1+\spll-\rho_O)^2+\sprp^2\right)}
               + \frac{3}{2} \arctan{\left(\frac{\rho_1+\spll-\rho_O}{\sprp}\right)} \,,
        \\
        \bullet \int\!\dif{\rho_1}\,  \rho_1~ \frac{(2(\rho_1+\spll-\rho_O)^2 + \sprp^2)\sprp}{\left(( \rho_1+\spll-\rho_O)^2+\sprp^2\right)^2}
            &= \frac{\sprp(\spll-\rho_O) (\rho_1 +\spll-\rho_O ) + \sprp^3}
                    {2 ((\rho_1 + (\spll-\rho_O))^2 + \sprp^2)  } 
               - \frac{3}{2} (\spll-\rho_O) \arctan\left(\frac{\rho_1 + \spll-\rho_O}{\sprp}\right)
            \\
            &\phantom{=} + \sprp \ln\left[(\rho_1 + \spll-\rho_O)^2 + \sprp^2\right]   \,, \nonumber
        \\
        \bullet \int\!\dif{\rho_1}\, \rho_1^2~ \frac{(2(\rho_1+\spll-\rho_O)^2 + \sprp^2)\sprp}{\left(( \rho_1+\spll-\rho_O)^2+\sprp^2\right)^2}
            &=   \frac{5}{2} \rho_1 \sprp
               - \frac{1}{2} (\spll-\rho_O) \sprp
               - \frac{\rho_1^2 \sprp(\rho_1+\spll - \rho_O)}
                      {2 ((\rho_1 + (\spll-\rho_O))^2 + \sprp^2)}
            \\ \nonumber
            &\phantom{=} + \frac{1}{2} \left(3(\spll-\rho_O)^2 - 5\sprp^2 \right) \arctan\left( \frac{\rho_1 + \spll - \rho_O} {\sprp} \right)
            \\ \nonumber
            &\phantom{=} - 2 (\spll-\rho_O) \sprp \ln\left[(\rho_1 + \spll-\rho_O)^2 + \sprp^2\right] \,,
        \\
        \bullet \int\!\dif{\rho_1}\, \rho_1^3~ \frac{(2(\rho_1+\spll-\rho_O)^2 + \sprp^2)\sprp}{\left(( \rho_1+\spll-\rho_O)^2+\sprp^2\right)^2} 
            &=  \frac{1}{2}\sprp \left( 3\rho_1^2  - 9 \rho_1 (\spll-\rho_O)  + (\spll-\rho_O)^2  - \sprp^2 \right)
            \\ \nonumber
            &\phantom{=} - \frac{\rho_1^3 \sprp (\rho_1+\spll - \rho_O)} {2 ((\rho_1 + \spll - \rho_O)^2 + \sprp^2)}
            \\ \nonumber
            &\phantom{=} - \frac{3}{2} (\spll-\rho_O) \left[(\spll-\rho_O)^2 - 5 \sprp^2\right] \arctan\left( \frac{\rho_1 + \spll - \rho_O}{\sprp} \right)
            \\ \nonumber
            &\phantom{=} + 3 \sprp  \left((\spll-\rho_O)^2 - \frac{1}{2} \sprp^2\right) \ln\left[(\rho_1 + (\spll-\rho_O))^2 + \sprp^2\right]\,.
    \end{align}
    We take particular care to ensure that $\rho_1=\rho_O-\spll$ is not one of
    the knots of the spline, because the integral is numerically unstable
    (though not analytically problematic) if an integral bound (knot point) is
    too close to this value.

    To do the full integral we need to know the range of $\spll$ and $\sprp$
    over which we should evaluate $\mcal{P}$.  We therefore identify a range
    $\rho_{\rm min}<\rho<\rho_{\rm max}$ over which $\GgamBar$ is above its
    value at $z_0$.  Requiring that
    \begin{equation}
        \rho_{\rm min} < \rho_1 < \rho_{\rm max}
    \end{equation}
    sets the range over which we integrate in \autoref{eq:Pspllsprpphi_app}
    while demanding that
    \begin{equation}
        \label{eq:spll_sprp_limits_app}
        \rho_{\rm min} < \rho_1 - \sqrt{(\rho_1+\spll-\rho_O)^2+ \sprp^2} < \rho_{\rm max}
    \end{equation}
    sets the range of $\spll$ and $\sprp$ as a function of $\rho_1$.  Rewriting
    the first inequality of \autoref{eq:spll_sprp_limits_app},
    \begin{equation}
        \sqrt{(\rho_1+\spll-\rho_O)^2+ \sprp^2} < \rho_1-\rho_{\rm min}
    \end{equation}
    and so
    \begin{equation} 
    \label{eqn:spllsprprange}
        (\rho_1+\spll-\rho_O)^2+ \sprp^2 < (\rho_1-\rho_{\rm min})^2 \,.
    \end{equation}
    Since the right hand side is largest when $\rho_1=\rho_{\rm max}$, we
    require
    \begin{equation}
        (\rho_1+\spll-\rho_O)^2+ \sprp^2 < (\rho_{\rm max}-\rho_{\rm min})^2 \,.
    \end{equation}
    We see that we must range over
    \begin{equation}
         0 \leq \sprp \leq \rho_{\rm max}-\rho_{\rm min}
    \end{equation}
    Rewriting \autoref{eqn:spllsprprange}, $\spll$ must be between the two roots
    $\sigma_\pm$ of the polynomial
    \begin{equation}
        P(\spll) \equiv
        \spll^2+2\spll(\rho_1-\rho_O) + (\rho_1-\rho_O)^2 + \sprp^2 - (\rho_1-\rho_{\rm min})^2
        \nonumber
    \end{equation}
    Thus
    \begin{equation}
        \sigma_\pm = 
         \rho_O-\rho_1 \pm 
         \sqrt{(\rho_1-\rho_{\rm min})^2 + \sprp^2}
    \end{equation}
    This translates to
    \begin{equation}
        \rho_O-\rho_{\rm min} \leq 
        \spll 
        \leq
        \rho_O-2\rho_{\rm max}+\rho_{\rm min}
        \leq 
        -\rho_O+\rho_{\rm min}
    \end{equation}
    For the obvious choice of $\rho_{\rm max}=\rho_O$, we get
    \begin{equation}
        -\rho_O + \rho_{\rm min} \leq \spll  \leq \rho_O - \rho_{\rm min} \,,
    \end{equation}
    which is nicely symmetric about $0$.

    Constant values of $\rho_2\leq\rho_1$ are circles in $(\spll,\sprp)$, with
    center $\rho_O-\rho_1$ and radius $\rho_1-\rho_2$.  Of course we can never
    have $\rho_2>\rho_1$.  We note that $\GgamBarCL(\rho)$ peaks at
    $\rho=\rho_R$ (recombination), so if $\rho_1<\rho_R$, then so is $\rho_2$,
    and $\GgamBarCL(\rho_2)$ never reaches its peak value.  However, if
    $\rho_1>\rho_R$, then $\GgamBarCL(\rho_2=\rho_R)$ is a circle of radius
    $\rho_1-\rho_R$, and center $\rho_O-\rho_1$.  These intersect $\sprp=0$ at
    $\spll=\rho_O-\rho_R$ and $\spll=\rho_O+\rho_R-2\rho_1$.

    \vspace{20pt}

\begin{figure}
        \script{p2ls/plot_p2ls_samples.py}
        \centering
        \includegraphics[width=1\columnwidth]{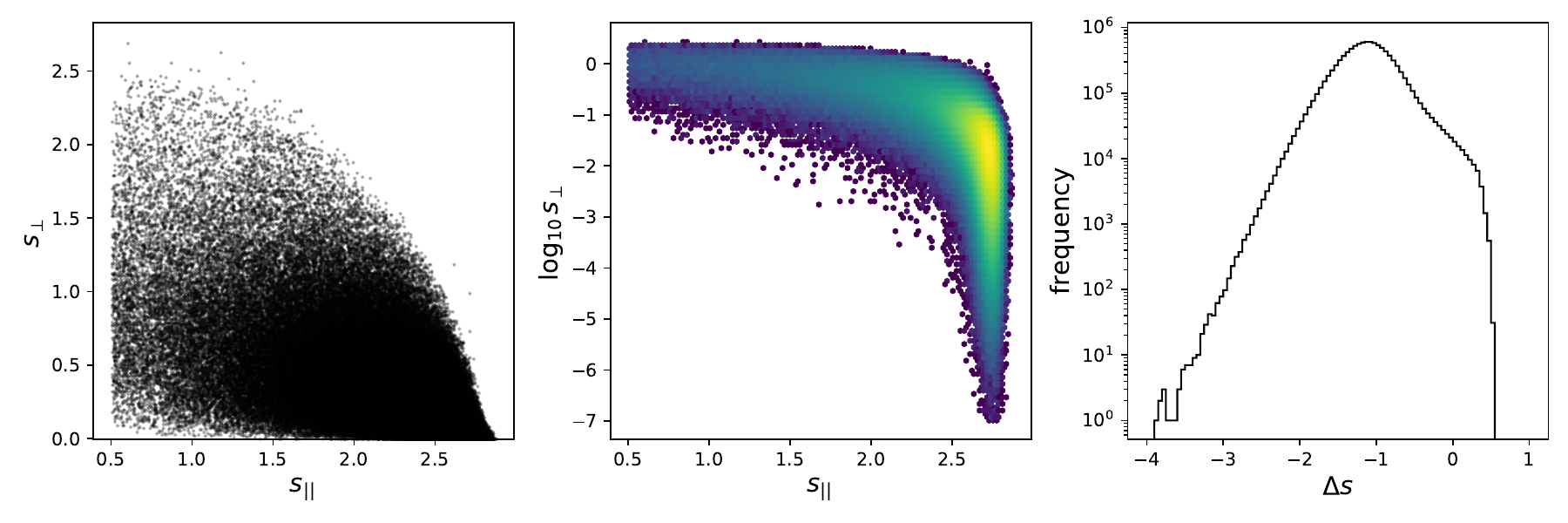}
        \caption{%
            10 million points sampled from $\mcal{P}$.  \textbf{Left:}
            $\mcal{P}$ the samples in $\spll, \sprp$.  The probability density
            function is negligible for $\spll < 0.5$, so we do not sample past
            this point.  \textbf{Middle:} same sample as \textbf{left}, plotting
            as a density histogram and in logarithmic coordinates for $\sprp$.
            The density coloring emphasizes that $\spll < 0.5$ is not important
            to the results.  The logarithmic coordinates shows that while
            $\sprp$ extends to 0 analytically, it is sampled linearly and thus
            decreases a factor of ten in sampled density for every decade in
            $\sprp \rightarrow 0$.  Our results are limited not by sampling
            $\sprp$ close to 0, but by samples of $\Delta s$ near 0.
            \textbf{Right:} showing the distribution of $\Delta s$.  The maximum
            separation is set by $(\spll, \sprp) = \norm{ (\rm{min}, \rm{max}) -
            (\rm{max}, \rm{min}) } \approx 3.2$.  The minimum separation is
            approximate $10^{-4}$, though may be decreased at the cost of
            sampling more points from $\mathcal{P}$.%
        }
        \label{fig:P2LS_samples}
    \end{figure}

    The Monte Carlo integration of \autoref{eq:signal_realspace} requires
    drawing samples from $\mcal{P}(\spll, \sprp, \phi)$, (defined in
    \autoref{eq:Pspllsprpphi} and  \autoref{eq:Pspllsprpphi_app}).  The $\phi$
    component is separable and can be trivially sampled from $\phi \sim U(0,
    2\pi)$.  The $\spll$ and $\sprp$ distributions are not separable.  To sample
    from $\mcal{P}(\spll, \sprp) \equiv 2\pi \mcal{P}(\spll, \sprp, \phi)$ we
    first sample $\spll$ from the marginal distribution distribution
    $\mcal{P}(\spll) = \int_{\sprp}\! \dif{\sprp} \,\mcal{P}(\spll, \sprp)$.
    Then we sample $\sprp$ from the conditional distribution $\mcal{P}(\sprp |
    \spll)$.  \autoref{fig:P2LS_samples} shows the set of sampled points used in
    \autoref{sec:calculating_the_signal} and
    \appref{app:calculating_the_correlation}.

% section calculating_and_sampling_P (end)

\section{Calculating the Correlation} \label{app:calculating_the_correlation}

    Calculating the correlation of $\Yd(\nhat_1)$ with $\Delta T^l(\nhat_2)$ is
    straightforward, if delicate, involving taking and simplifying many
    expectation values.  We do not leave this as an exercise to the reader.
    \begin{align}
        \frac{1}{T_0^2}\Yd(\nhat_1)(\DeltaT^l(\nhat_2))^2
            &= \frac{1}{2 T_0^4} \Bigg[
                \int\!\!\! \dif{\mbf{s}_1} \; \mcal{P}(\mbf{s}_1)  \left(\DeltaT^e(\mbf{x}(\nhat_1,\mbf{s}_1))\right)^2
                - \int\!\! \dif{\mbf{s}_1} \mcal{P}(\mbf{s}_1) \dif{\mbf{s}'_1} \mcal{P}(\mbf{s}'_1) \DeltaT^e(\mbf{x}(\nhat_1,\mbf{s}_1))\DeltaT^e(\mbf{x}(\nhat_1,\mbf{s}'_1)) \Bigg]
            \\
            &\qquad \times \int\!\! \dif{\mbf{s}_2} \mcal{P}(\mbf{s}_2) \dif{\mbf{s}'_2} \mcal{P}(\mbf{s}'_2) 
            \DeltaT^l(\mbf{x}(\nhat_2,\mbf{s}_2))\DeltaT^l(\mbf{x}(\nhat_2,\mbf{s}'_2))
         \nonumber
        \\
            &= \frac{1}{2 T_0^4} \Bigg[
                \phantom{+} \int\!\! \dif{\mbf{s}_1} \mcal{P}(\mbf{s}_1) \dif{\mbf{s}_2} \mcal{P}(\mbf{s}_2) \dif{\mbf{s}'_2} \mcal{P}(\mbf{s}'_2) \left(\DeltaT^e(\mbf{x}(\nhat_1,\mbf{s}_1))\right)^2 \DeltaT^l(\mbf{x}(\nhat_2,\mbf{s}_2)) \DeltaT^l(\mbf{x}(\nhat_2,\mbf{s}'_2))
                \\ \nonumber & \qquad\quad
                - \int\!\! \dif{\mbf{s}_1} \mcal{P}(\mbf{s}_1) \dif{\mbf{s}'_1} \mcal{P}(\mbf{s}'_1) \dif{\mbf{s}_2} \mcal{P}(\mbf{s}_2) \dif{\mbf{s}'_2} \mcal{P}(\mbf{s}'_2) \DeltaT^e(\mbf{x}(\nhat_1,\mbf{s}_1))\DeltaT^e(\mbf{x}(\nhat_1,\mbf{s}'_1)) \DeltaT^l(\mbf{x}(\nhat_2,\mbf{s}_2)) \DeltaT^l(\mbf{x}(\nhat_2,\mbf{s}'_2))
                \Bigg] \nonumber \,.
        \intertext{Taking the expectation value,}
        2 T_0^4 \langle \frac{1}{T_0^2}\Yd(\nhat_1)(\DeltaT^l(\nhat_2))^2 \rangle
            &= 
                \phantom{+} \int\!\! \dif{\mbf{s}_1} \mcal{P}(\mbf{s}_1) \dif{\mbf{s}_2} \mcal{P}(\mbf{s}_2) \dif{\mbf{s}'_2} \mcal{P}(\mbf{s}'_2) \Big\langle \left(\DeltaT^e(\mbf{x}(\nhat_1,\mbf{s}_1)\right)^2) \DeltaT^l(\mbf{x}(\nhat_2,\mbf{s}_2)) \DeltaT^l(\mbf{x}(\nhat_2,\mbf{s}'_2)) \Big\rangle
                \\ \nonumber & \quad
                - \int\!\! \dif{\mbf{s}_1} \mcal{P}(\mbf{s}_1) \dif{\mbf{s}'_1} \mcal{P}(\mbf{s}'_1) \dif{\mbf{s}_2} \mcal{P}(\mbf{s}_2) \dif{\mbf{s}'_2} \mcal{P}(\mbf{s}'_2) \Big\langle \DeltaT^e(\mbf{x}(\nhat_1,\mbf{s}_1)) \DeltaT^e(\mbf{x}(\nhat_1,\mbf{s}'_1)) \DeltaT^l(\mbf{x}(\nhat_2,\mbf{s}_2)\Delta) T^l(\mbf{x}(\nhat_2,\mbf{s}'_2) \Big\rangle
            \\
            &=   \nonumber
                \phantom{+} \int\!\! \dif{\mbf{s}_1} \mcal{P}(\mbf{s}_1) \dif{\mbf{s}_2} \mcal{P}(\mbf{s}_2) \dif{\mbf{s}'_2} \mcal{P}(\mbf{s}'_2)
                    \Big\langle \left(\DeltaT^e(\mbf{x}(\nhat_1,\mbf{s}_1)\right)^2) \Big\rangle
                    \Big\langle \DeltaT^l(\mbf{x}(\nhat_2,\mbf{s}_2)) \DeltaT^l(\mbf{x}(\nhat_2,\mbf{s}'_2)) \Big\rangle
                \\ & \quad
                + 2 \int\!\! \dif{\mbf{s}_1} \mcal{P}(\mbf{s}_1) \dif{\mbf{s}_2} \mcal{P}(\mbf{s}_2) \dif{\mbf{s}'_2} \mcal{P}(\mbf{s}'_2)
                    \Big\langle \DeltaT^e(\mbf{x}(\nhat_1,\mbf{s}_1)) \DeltaT^l(\mbf{x}(\nhat_2,\mbf{s}_2)) \Big\rangle
                    \Big\langle \DeltaT^e(\mbf{x}(\nhat_1,\mbf{s}_1)) \DeltaT^l(\mbf{x}(\nhat_2,\mbf{s}'_2)) \Big\rangle
                \nonumber \\ & \quad
                - \int\!\! \dif{\mbf{s}_1} \mcal{P}(\mbf{s}_1) \dif{\mbf{s}'_1} \mcal{P}(\mbf{s}'_1) \dif{\mbf{s}_2} \mcal{P}(\mbf{s}_2) \dif{\mbf{s}'_2} \mcal{P}(\mbf{s}'_2)
                    \Bigg[
                    \phantom{+}
                    \Big\langle \DeltaT^e(\mbf{x}(\nhat_1,\mbf{s}_1)) \DeltaT^e(\mbf{x}(\nhat_1,\mbf{s}'_1)) \Big\rangle
                    \Big\langle \DeltaT^l(\mbf{x}(\nhat_2,\mbf{s}_2)) \DeltaT^l(\mbf{x}(\nhat_2,\mbf{s}'_2)) \Big\rangle
                    \nonumber \\ & \qquad\qquad\qquad\qquad\qquad\qquad\qquad\qquad\qquad\quad
                    +
                    \Big\langle \DeltaT^e(\mbf{x}(\nhat_1,\mbf{s}_1)) \DeltaT^l(\mbf{x}(\nhat_2,\mbf{s}_2)) \Big\rangle
                    \Big\langle \DeltaT^e(\mbf{x}(\nhat_1,\mbf{s}'_1)) \DeltaT^l(\mbf{x}(\nhat_2,\mbf{s}'_2)) \Big\rangle
                     \nonumber \\ & \qquad\qquad\qquad\qquad\qquad\qquad\qquad\qquad\qquad\quad
                    +
                    \Big\langle \DeltaT^e(\mbf{x}(\nhat_1,\mbf{s}_1)) \DeltaT^l(\mbf{x}(\nhat_2,\mbf{s}'_2)) \Big\rangle
                    \Big\langle \DeltaT^e(\mbf{x}(\nhat_1,\mbf{s}'_1)) \DeltaT^l(\mbf{x}(\nhat_2,\mbf{s}_2)) \Big\rangle
                \Bigg] \nonumber
            \\
            &=  \nonumber
                \phantom{+} \int\!\! \dif{\mbf{s}_2} \mcal{P}(\mbf{s}_2) \dif{\mbf{s}'_2} \mcal{P}(\mbf{s}'_2)
                    \xi^{ee}(0)
                    \xi^{ll}\left(\norm{ \mbf{x}(\nhat_2,\mbf{s}_2) - \mbf{x}(\nhat_2,\mbf{s}'_2) }\right) 
                    \label{eq:yTTexpecation_value}
                \\ & \quad
                + 2 \int\!\! \dif{\mbf{s}_1} \mcal{P}(\mbf{s}_1) \dif{\mbf{s}_2} \mcal{P}(\mbf{s}_2) \dif{\mbf{s}'_2} \mcal{P}(\mbf{s}'_2)
                    \xi^{el}\left(\norm{ \mbf{x}(\nhat_1,\mbf{s}_1) - \mbf{x}(\nhat_2,\mbf{s}_2) }\right)
                    \xi^{el}\left(\norm{ \mbf{x}(\nhat_1,\mbf{s}_1) - \mbf{x}(\nhat_2,\mbf{s}'_2) }\right)
                \nonumber \\ & \quad
                - \int\!\! \dif{\mbf{s}_1} \mcal{P}(\mbf{s}_1) \dif{\mbf{s}'_1} \mcal{P}(\mbf{s}'_1) \dif{\mbf{s}_2} \mcal{P}(\mbf{s}_2) \dif{\mbf{s}'_2} \mcal{P}(\mbf{s}'_2)
                    \Bigg[
                    \phantom{+}
                    \xi^{ee}\left(\norm{ \mbf{x}(\nhat_1,\mbf{s}_1) - \mbf{x}(\nhat_1,\mbf{s}'_1) }\right)
                    \xi^{ll}\left(\norm{ \mbf{x}(\nhat_2,\mbf{s}_2) - \mbf{x}(\nhat_2,\mbf{s}'_2) }\right)
                    \nonumber \\ & \qquad\qquad\qquad\qquad\qquad\qquad\qquad\qquad\qquad\quad
                    +
                    \xi^{el}\left(\norm{ \mbf{x}(\nhat_1,\mbf{s}_1) - \mbf{x}(\nhat_2,\mbf{s}_2) }\right)
                    \xi^{el}\left(\norm{ \mbf{x}(\nhat_1,\mbf{s}'_1) - \mbf{x}(\nhat_2,\mbf{s}'_2) }\right)
                     \nonumber \\ & \qquad\qquad\qquad\qquad\qquad\qquad\qquad\qquad\qquad\quad
                    +
                    \xi^{el}\left(\norm{ \mbf{x}(\nhat_1,\mbf{s}_1) - \mbf{x}(\nhat_2,\mbf{s}'_2) }\right)
                    \xi^{el}\left(\norm{ \mbf{x}(\nhat_1,\mbf{s}'_1) - \mbf{x}(\nhat_2,\mbf{s}_2) }\right)
                \Bigg] \nonumber \,.
        \intertext{The correlation is written in terms of \autoref{eq:yTTexpecation_value}, where the subtraction of the individual expectation values $\Big\langle\Yd(\nhat_1)\Big\rangle$ and $\Big\langle(\DeltaT(\nhat_2))^2 \Big\rangle$ greatly simplifies the final equation.}
        C^{\Yd(\DeltaT)^2}(\mbf{\nhat}_1,\mbf{\nhat}_2) 
            & \equiv
                \Big\langle \frac{1}{T_0^2}\Yd(\nhat_1)(\DeltaT(\nhat_2))^2 \Big\rangle - \frac{1}{T_0^2} \Big\langle\Yd(\nhat_1)\Big\rangle \Big\langle(\DeltaT(\nhat_2))^2 \Big\rangle 
            \\
            &= 
                \phantom{+}\,\frac{1}{T_0^4} \int\!\! \dif{\mbf{s}_1} \mcal{P}(\mbf{s}_1) \dif{\mbf{s}_2} \mcal{P}(\mbf{s}_2) \dif{\mbf{s}'_2} \mcal{P}(\mbf{s}'_2)
                    \xi^{el}\left(\norm{ \mbf{x}(\nhat_1,\mbf{s}_1) - \mbf{x}(\nhat_2,\mbf{s}_2) }\right)
                    \xi^{el}\left(\norm{ \mbf{x}(\nhat_1,\mbf{s}_1) - \mbf{x}(\nhat_2,\mbf{s}'_2) }\right)
                \nonumber \\ & \quad
                - \frac{1}{2 T_0^4} \int\!\! \dif{\mbf{s}_1} \mcal{P}(\mbf{s}_1) \dif{\mbf{s}'_1} \mcal{P}(\mbf{s}'_1) \dif{\mbf{s}_2} \mcal{P}(\mbf{s}_2) \dif{\mbf{s}'_2} \mcal{P}(\mbf{s}'_2)
                    \Bigg[
                    \xi^{el}\left(\norm{ \mbf{x}(\nhat_1,\mbf{s}_1) - \mbf{x}(\nhat_2,\mbf{s}_2) }\right)
                    \xi^{el}\left(\norm{ \mbf{x}(\nhat_1,\mbf{s}'_1) - \mbf{x}(\nhat_2,\mbf{s}'_2) }\right)
                     \nonumber \\ & \qquad\qquad\qquad\qquad\qquad\qquad\qquad\qquad\qquad\qquad\quad
                    +
                    \xi^{el}\left(\norm{ \mbf{x}(\nhat_1,\mbf{s}_1) - \mbf{x}(\nhat_2,\mbf{s}'_2) }\right)
                    \xi^{el}\left(\norm{ \mbf{x}(\nhat_1,\mbf{s}'_1) - \mbf{x}(\nhat_2,\mbf{s}_2) }\right)
                \Bigg] \nonumber
            \\
            &= 
                \phantom{+}\frac{1}{T_0^4} \int\!\! \dif{\mbf{s}_1} \mcal{P}(\mbf{s}_1) \dif{\mbf{s}_2} \mcal{P}(\mbf{s}_2) \dif{\mbf{s}'_2} \mcal{P}(\mbf{s}'_2)
                    \xi^{el}\left(\norm{ \mbf{x}(\nhat_1,\mbf{s}_1) - \mbf{x}(\nhat_2,\mbf{s}_2) }\right)
                    \xi^{el}\left(\norm{ \mbf{x}(\nhat_1,\mbf{s}_1) - \mbf{x}(\nhat_2,\mbf{s}'_2) }\right)
                \nonumber \\ & \quad
                - \frac{1}{T_0^4} \left( \int\!\! \dif{\mbf{s}_1} \mcal{P}(\mbf{s}_1) \dif{\mbf{s}_2} \mcal{P}(\mbf{s}_2)
                    \xi^{el}\left(\norm{ \mbf{x}(\nhat_1,\mbf{s}_1) - \mbf{x}(\nhat_2,\mbf{s}_2) }\right)
                    \right)^2 \,.
                \nonumber
    \end{align}

% section calculating_the_correlation (end)

%%%%%%%%%%%%%%%%%%%%%%%%%%%%%%%%%%%%%%%%%%%%%%%%%%%%%%%%%%%%%%%%%%%%%%%%%%%%%%%

\label{lastpage}

\end{document}

%% file: preamble.tex
\newcommand{\appref}[1]{\hyperref[#1]{Appendix~\autoref*{#1}}}

\newcommand{\norm}[1]{\lvert #1 \rvert}
\newcommand{\mbf}[1]{\mathbf{#1}}
\newcommand{\mcal}[1]{\mathcal{#1}}

\newcommand{\Exp}[1]{e^{#1}}

\NewDocumentCommand \dif {o m} {
	\IfNoValueTF{#1}
		{ \mathrm{d}{#2} }
		{ \mathrm{d}^{#1}{\!#2} }
}

\NewDocumentCommand \deriv {s O{x} d<> m} {
	\IfBooleanTF{#1}
		{% 
		 \IfNoValueTF{#3}
		 { \frac{\mathrm{d}}{\dif{#2}}\left({#4}\right) }
		 { \frac{\mathrm{d}^{#3}}{\dif{#2}^{#3}}\left({#4}\right) }
		}
		{%
		 \IfNoValueTF{#3}
		 { \frac{\mathrm{d}{#4}}{\dif{#2}} }
		 { \frac{\dif[{#3}]{#4}}{\dif{#2}^{#3}} }
		}
}

\newcommand{\Leq}{L_{\rm eq}}
\newcommand{\spll}{s_{||}}
\newcommand{\sprp}{s_{\perp}}
\newcommand{\sreco}{s_{\rm recombination}}
\newcommand{\nhat}{{\bf\hat{n}}}

\newcommand{\Pgam}{P_{\gamma}}
\newcommand{\PgamBar}{\overline{\Pgam}}
\newcommand{\PgamBarCL}{\PgamBar^{CL}}

\newcommand{\Ggam}{g_\gamma}
\newcommand{\GgamBar}{\overline{\Ggam}}
\newcommand{\GgamBarCL}{\GgamBar^{CL}}

\newcommand{\aeq}{a_{\rm eq}}
\newcommand{\OmO}{\Omega_{m,0}}

\newcommand{\DeltaT}{{\Delta T}}
\newcommand{\Yd}{\mathcal{Y}}

\newcommand{\package}[1]{\textsc{#1}}
\newcommand{\LCDM}{$\Lambda$CDM}

%% file: graphic.tex
\begin{tikzpicture}[x=0.75pt,y=0.75pt,yscale=-1,xscale=1]

%Straight Lines [id:da813715462089712] 
\draw  [dash pattern={on 4.5pt off 4.5pt}]  (98.01,106.04) -- (587,105) ;
\draw [shift={(95.01,106.04)}, rotate = 359.88] [fill={rgb, 255:red, 0; green, 0; blue, 0 }  ][line width=0.08]  [draw opacity=0] (8.93,-4.29) -- (0,0) -- (8.93,4.29) -- cycle    ;

%Straight Lines [id:da2552912430294406] 
\draw [color={rgb, 255:red, 208; green, 2; blue, 27 }  ,draw opacity=1 ] [dash pattern={on 4.5pt off 4.5pt}]  (104.33,131) -- (596,130.01) ;
\draw [shift={(599,130)}, rotate = 179.88] [fill={rgb, 255:red, 208; green, 2; blue, 27 }  ,fill opacity=1 ][line width=0.08]  [draw opacity=0] (8.93,-4.29) -- (0,0) -- (8.93,4.29) -- cycle    ;

%Straight Lines [id:da35530834229993047] 
\draw [color={rgb, 255:red, 208; green, 2; blue, 27 }  ,draw opacity=1 ]   (171.33,119.67) -- (171.33,141) ;

%Straight Lines [id:da021420636439809182] 
\draw [color={rgb, 255:red, 208; green, 2; blue, 27 }  ,draw opacity=1 ]   (281.33,119.67) -- (281.33,141) ;

%Straight Lines [id:da6128001501161549] 
\draw [color={rgb, 255:red, 208; green, 2; blue, 27 }  ,draw opacity=1 ]   (492.33,119.67) -- (492.33,141) ;

%Shape: Brace [id:dp6543544919744094] 
\draw  [color={rgb, 255:red, 128; green, 128; blue, 128 }  ,draw opacity=1 ] (491.06,85.03) .. controls (491.06,80.36) and (488.73,78.03) .. (484.06,78.03) -- (396.06,78.03) .. controls (389.39,78.03) and (386.06,75.7) .. (386.06,71.03) .. controls (386.06,75.7) and (382.73,78.03) .. (376.06,78.03)(379.06,78.03) -- (288.06,78.03) .. controls (283.39,78.03) and (281.06,80.36) .. (281.06,85.03) ;

%Image [id:dp4881354627714134] 
\draw (599.5,102.67) node  {\includegraphics[width=18.75pt,height=19pt]{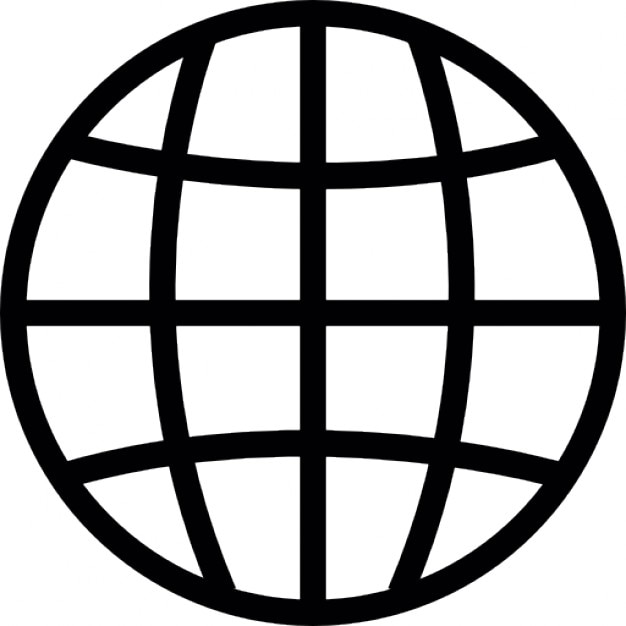}};
%Straight Lines [id:da1711541559215889] 
\draw [color={rgb, 255:red, 221; green, 206; blue, 35 }  ,draw opacity=1 ]   (281.67,106.33) .. controls (283.32,104.66) and (284.99,104.65) .. (286.67,106.3) .. controls (288.34,107.95) and (290.01,107.94) .. (291.67,106.27) .. controls (293.32,104.59) and (294.99,104.58) .. (296.67,106.23) .. controls (298.34,107.88) and (300.01,107.87) .. (301.67,106.2) .. controls (303.33,104.53) and (305,104.52) .. (306.67,106.17) .. controls (308.34,107.82) and (310.01,107.81) .. (311.67,106.14) .. controls (313.32,104.46) and (314.99,104.45) .. (316.67,106.1) .. controls (318.34,107.75) and (320.01,107.74) .. (321.67,106.07) .. controls (323.33,104.4) and (325,104.39) .. (326.67,106.04) .. controls (328.35,107.69) and (330.02,107.68) .. (331.67,106) .. controls (333.33,104.33) and (335,104.32) .. (336.67,105.97) .. controls (338.34,107.62) and (340.01,107.61) .. (341.67,105.94) .. controls (343.33,104.27) and (345,104.26) .. (346.67,105.91) .. controls (348.35,107.56) and (350.02,107.55) .. (351.67,105.87) .. controls (353.33,104.2) and (355,104.19) .. (356.67,105.84) .. controls (358.34,107.49) and (360.01,107.48) .. (361.66,105.81) .. controls (363.31,104.13) and (364.98,104.12) .. (366.66,105.77) .. controls (368.33,107.42) and (370,107.41) .. (371.66,105.74) .. controls (373.32,104.07) and (374.99,104.06) .. (376.66,105.71) .. controls (378.33,107.36) and (380,107.35) .. (381.66,105.68) .. controls (383.31,104) and (384.98,103.99) .. (386.66,105.64) .. controls (388.33,107.29) and (390,107.28) .. (391.66,105.61) .. controls (393.32,103.94) and (394.99,103.93) .. (396.66,105.58) .. controls (398.34,107.23) and (400.01,107.22) .. (401.66,105.54) .. controls (403.32,103.87) and (404.99,103.86) .. (406.66,105.51) .. controls (408.33,107.16) and (410,107.15) .. (411.66,105.48) .. controls (413.31,103.8) and (414.98,103.79) .. (416.66,105.44) .. controls (418.33,107.09) and (420,107.08) .. (421.66,105.41) .. controls (423.32,103.74) and (424.99,103.73) .. (426.66,105.38) .. controls (428.33,107.03) and (430,107.02) .. (431.66,105.35) .. controls (433.31,103.67) and (434.98,103.66) .. (436.66,105.31) .. controls (438.33,106.96) and (440,106.95) .. (441.66,105.28) .. controls (443.32,103.61) and (444.99,103.6) .. (446.66,105.25) .. controls (448.34,106.9) and (450.01,106.89) .. (451.66,105.21) .. controls (453.32,103.54) and (454.99,103.53) .. (456.66,105.18) .. controls (458.33,106.83) and (460,106.82) .. (461.66,105.15) .. controls (463.32,103.48) and (464.99,103.47) .. (466.66,105.12) .. controls (468.34,106.77) and (470.01,106.76) .. (471.66,105.08) .. controls (473.32,103.41) and (474.99,103.4) .. (476.66,105.05) -- (481.63,105.02) -- (489.63,104.96) ;
\draw [shift={(492.63,104.95)}, rotate = 179.62] [fill={rgb, 255:red, 221; green, 206; blue, 35 }  ,fill opacity=1 ][line width=0.08]  [draw opacity=0] (8.93,-4.29) -- (0,0) -- (8.93,4.29) -- cycle    ;
\draw  [color={rgb, 255:red, 128; green, 128; blue, 128 }  ,draw opacity=1 ] (281.5,85) .. controls (283.29,80.69) and (282.04,77.64) .. (277.73,75.85) -- (249.94,64.29) .. controls (243.78,61.72) and (241.6,58.29) .. (243.39,53.98) .. controls (241.6,58.29) and (237.62,59.16) .. (231.47,56.6)(234.24,57.76) -- (203.68,45.04) .. controls (199.37,43.25) and (196.32,44.51) .. (194.53,48.82) ;

%Shape: Arc [id:dp2490690148764544] 
\draw  [draw opacity=0] (253.24,105.22) .. controls (253.24,105.22) and (253.24,105.22) .. (253.24,105.22) .. controls (252.76,101.52) and (254.14,98.05) .. (256.6,95.89) -- (262.67,104) -- cycle ; \draw  [color={rgb, 255:red, 128; green, 128; blue, 128 }  ,draw opacity=1 ] (253.24,105.22) .. controls (253.24,105.22) and (253.24,105.22) .. (253.24,105.22) .. controls (252.76,101.52) and (254.14,98.05) .. (256.6,95.89) ; 

%Straight Lines [id:da5578233436423486] 
\draw [color={rgb, 255:red, 221; green, 206; blue, 35 }  ,draw opacity=1 ]  (184.61,67.86) .. controls (185.55,70.02) and (187.1,70.63) .. (189.26,69.7) .. controls (191.43,68.77) and (192.98,69.38) .. (193.91,71.55) .. controls (194.85,73.71) and (196.4,74.32) .. (198.56,73.39) .. controls (200.72,72.46) and (202.27,73.07) .. (203.2,75.23) .. controls (204.14,77.39) and (205.69,78) .. (207.85,77.07) .. controls (210.02,76.14) and (211.57,76.75) .. (212.5,78.92) .. controls (213.44,81.08) and (214.99,81.69) .. (217.15,80.76) .. controls (219.31,79.83) and (220.86,80.44) .. (221.8,82.6) .. controls (222.73,84.76) and (224.28,85.37) .. (226.44,84.44) .. controls (228.61,83.51) and (230.16,84.12) .. (231.09,86.29) .. controls (232.03,88.45) and (233.58,89.06) .. (235.74,88.13) .. controls (237.9,87.2) and (239.45,87.81) .. (240.39,89.97) .. controls (241.33,92.13) and (242.88,92.74) .. (245.04,91.81) .. controls (247.21,90.88) and (248.76,91.49) .. (249.69,93.66) .. controls (250.62,95.82) and (252.17,96.43) .. (254.33,95.5) .. controls (256.49,94.57) and (258.04,95.18) .. (258.98,97.34) .. controls (259.92,99.5) and (261.47,100.11) .. (263.63,99.18) .. controls (265.8,98.25) and (267.35,98.86) .. (268.28,101.03) -- (271.44,102.28) -- (278.88,105.23) ;
\draw [shift={(281.67,106.33)}, rotate = 201.62] [fill={rgb, 255:red, 221; green, 206; blue, 35 }  ,fill opacity=1 ][line width=0.08]  [draw opacity=0] (8.93,-4.29) -- (0,0) -- (8.93,4.29) -- cycle    ;
%Straight Lines [id:da39503913052661443] 
\draw    (171.33,117) -- (171.33,95.33) ;
%Straight Lines [id:da4051172508030588] 
\draw    (281.67,117) -- (281.67,95.33) ;
%Straight Lines [id:da868662886573279] 
\draw    (492.33,117) -- (492.33,95.33) ;
%Image [id:dp1131827792049599] 
\draw (493,92) node  {\includegraphics[width=16.5pt,height=14pt]{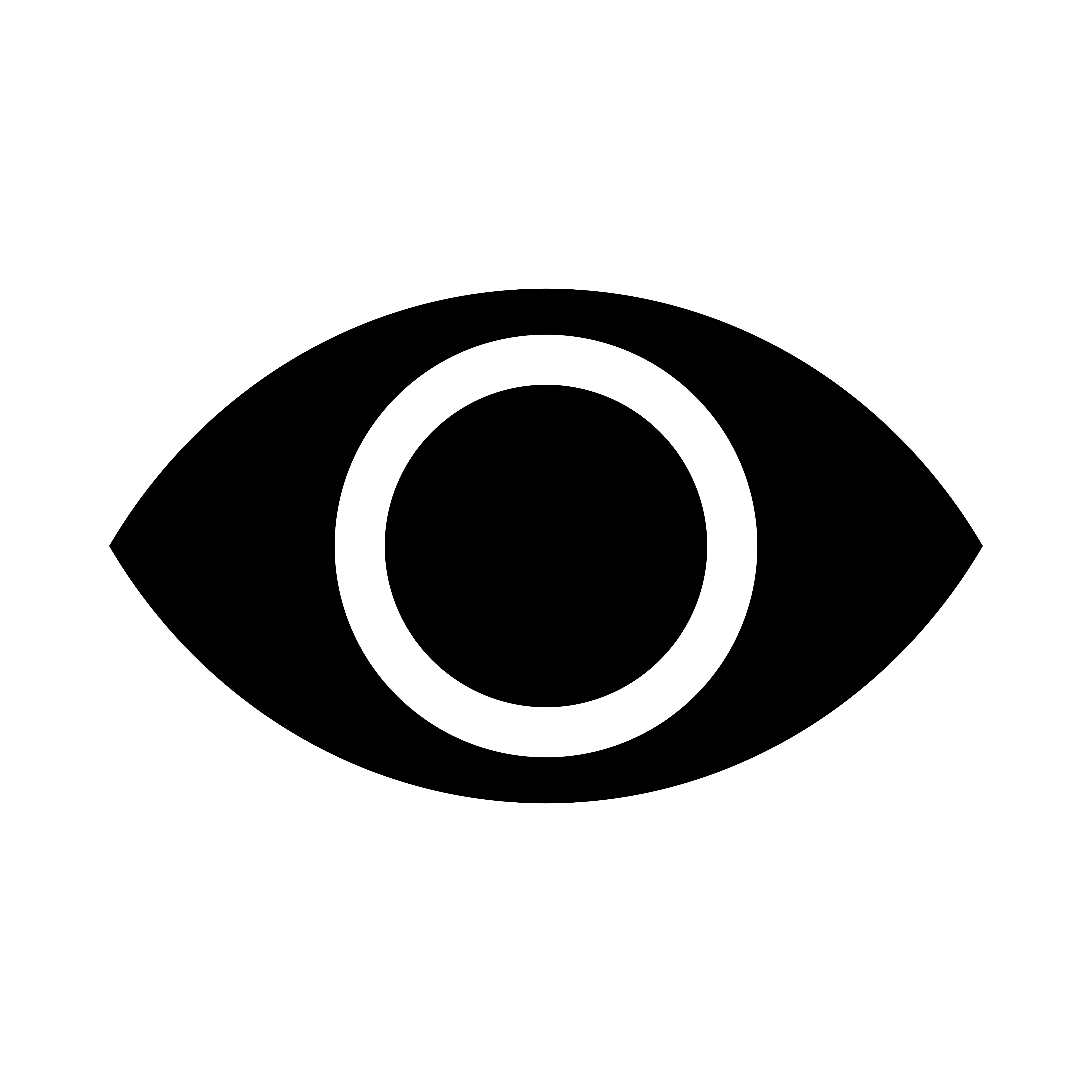}};
%Shape: Arc [id:dp5409355552805143] 
\draw  [draw opacity=0] (281.67,107.33) .. controls (282.41,113.47) and (285.17,118.02) .. (288.47,118.02) .. controls (292.34,118.02) and (295.47,111.76) .. (295.47,104.02) .. controls (295.47,96.29) and (292.34,90.02) .. (288.47,90.02) .. controls (285.5,90.02) and (282.95,93.73) .. (281.94,98.97) -- (288.47,104.02) -- cycle ; \draw [color={rgb, 255:red, 128; green, 128; blue, 128 }  ,draw opacity=1 ]   (282.21,110.31) .. controls (283.37,114.89) and (285.73,118.02) .. (288.47,118.02) .. controls (292.34,118.02) and (295.47,111.76) .. (295.47,104.02) .. controls (295.47,96.29) and (292.34,90.02) .. (288.47,90.02) .. controls (285.5,90.02) and (282.95,93.73) .. (281.94,98.97) ;  \draw [shift={(281.67,107.33)}, rotate = 66.78] [fill={rgb, 255:red, 128; green, 128; blue, 128 }  ,fill opacity=1 ][line width=0.08]  [draw opacity=0] (8.93,-4.29) -- (0,0) -- (8.93,4.29) -- cycle    ;

%Straight Lines [id:da3349166445996694] 
\draw [color={rgb, 255:red, 128; green, 128; blue, 128 }  ,draw opacity=1 ]   (492.63,104.95) -- (496.88,107.78) -- (508.17,114) ;
%Shape: Rectangle [id:dp5447073527145074] 
\draw  [color={rgb, 255:red, 128; green, 128; blue, 128 }  ,draw opacity=1 ] (492.63,104.95) -- (498.16,108.32) -- (498.17,116.5) -- (492.64,112.63) -- cycle ;

%Straight Lines [id:da31628460956113036] 
\draw [color={rgb, 255:red, 128; green, 128; blue, 128 }  ,draw opacity=1 ]   (449.63,104.95) -- (492.63,104.95) ;
\draw [shift={(446.63,104.95)}, rotate = 0] [fill={rgb, 255:red, 128; green, 128; blue, 128 }  ,fill opacity=1 ][line width=0.08]  [draw opacity=0] (8.93,-4.29) -- (0,0) -- (8.93,4.29) -- cycle    ;

%Shape: Arc [id:dp7660617208397223] 
\draw  [draw opacity=0] (172.06,106.89) .. controls (172.02,106.14) and (172,105.39) .. (172,104.64) .. controls (171.96,90.84) and (177.6,78.31) .. (186.77,69.18) -- (223.75,104.5) -- cycle ; \draw [color={rgb, 255:red, 128; green, 128; blue, 128 }  ,draw opacity=1 ]  (172.06,106.89) .. controls (172.02,106.14) and (172,105.39) .. (172,104.64) .. controls (171.96,90.84) and (177.6,78.31) .. (186.77,69.18) ;

% Text Node
\draw (220,24) node [anchor=north west][inner sep=0.75pt]  [color={rgb, 255:red, 128; green, 128; blue, 128 }  ,opacity=1 ,rotate=-24.18]  {$r_{12} =|\mathbf{x}_{2} -\mathbf{x}_{1} |$};
% Text Node
\draw (164,140) node [anchor=north west][inner sep=0.75pt]  [color={rgb, 255:red, 208; green, 2; blue, 27 }  ,opacity=1 ]  {$z_{2}$};
% Text Node
\draw (274,140) node [anchor=north west][inner sep=0.75pt]  [color={rgb, 255:red, 208; green, 2; blue, 27 }  ,opacity=1 ]  {$z_{1}$};
% Text Node
\draw (485,140) node [anchor=north west][inner sep=0.75pt]  [color={rgb, 255:red, 208; green, 2; blue, 27 }  ,opacity=1 ]  {$z_{0}$};
% Text Node
\draw (593,134.4) node [anchor=north west][inner sep=0.75pt]  [color={rgb, 255:red, 208; green, 2; blue, 27 }  ,opacity=1 ]  {$z$};
% Text Node
\draw (97.01,111.44) node [anchor=north west][inner sep=0.75pt]    {$\hat{n}$};
% Text Node
\draw (243,95) node [anchor=north west][inner sep=0.75pt]  [color={rgb, 255:red, 128; green, 128; blue, 128 }  ,opacity=1 ]  {$\theta $};
% Text Node
\draw (352,55) node [anchor=north west][inner sep=0.75pt]  [color={rgb, 255:red, 128; green, 128; blue, 128 }  ,opacity=1 ]  {$r_{01} =|\mathbf{x}_{1} -\mathbf{x}_{0} |$};
% Text Node
\draw (269,108) node [anchor=north west][inner sep=0.75pt]  [color={rgb, 255:red, 128; green, 128; blue, 128 }  ,opacity=1 ]  {$\phi $};

% Text Node
\draw (498.88,115) node [anchor=north west][inner sep=0.75pt]  [color={rgb, 255:red, 128; green, 128; blue, 128 }  ,opacity=1 ]  {$s_{\perp }$};
% Text Node
\draw (457,107) node [anchor=north west][inner sep=0.75pt]  [color={rgb, 255:red, 128; green, 128; blue, 128 }  ,opacity=1 ]  {$s_{||}\hat{n}$};

\end{tikzpicture}

%% file: output/signal.txt
9.0 \times 10^{-11}

%% file: output/undamped_signal.txt
2.8 \times 10^{-9}

%% file: output/signal_ratio.txt
31

%% file: output/sn_act.txt
12